\documentclass[preprint,10pt]{elsarticle}
\usepackage{geometry}
\usepackage{pdflscape}
\usepackage[table]{xcolor}
\usepackage{array}
\usepackage{rotating}
\usepackage{amssymb}
\usepackage{amsmath}
\usepackage{tikz}
\usepackage{amsmath}
\usepackage{booktabs}
\usepackage[shortlabels]{enumitem}
\usepackage{multirow}
\usepackage{lipsum}
\usepackage{lineno}
\usepackage{multicol}
\usepackage{enumitem}
\usepackage{rotating}
\usepackage{comment}
\usepackage{pifont}
\usepackage{tabularx}
\usepackage{array}

\usepackage{array}      
\usepackage{tabularx}   
\usepackage{booktabs}   

\newcommand{\revfirstremove}[1]{}
\newcommand{\revsecondremove}[1]{}
\newcolumntype{Y}{>{\arraybackslash}X}

\usepackage[acronym,nomain]{glossaries}
\glsdisablehyper

\newacronym{sta}{SO}{Stack Overflow}
\newacronym{sst}{SST}{Software Security Testing}
\newacronym{cicd}{CI/CD}{Continuous Integration/Continuous Deployment}
\newacronym{orm}{ORM}{Object Relationship Model}
\newacronym{csp}{CSP}{Content Security Policy}
\newacronym{sast}{SAST}{Static Application Security Testing}
\newacronym{dast}{DAST}{Dynamic Application Security Testing}
\newacronym{iast}{IAST}{Interactive Application Security Testing}
\newacronym{sdlc}{SDLC}{Software Development Life Cycle}
\newacronym{softeng}{SE}{Software Engineering}

\usepackage{url}

\usepackage{breakurl}

\usepackage{pgfplots}
\pgfplotsset{compat=1.18}

\usepackage[breaklinks]{hyperref}

\usepackage{cleveref}
\crefname{section}{§}{§§}
\Crefname{section}{§}{§§}

\usepackage{booktabs}
\usepackage{array}

\usepackage{longtable}
\usepackage{xcolor}
\usepackage[most]{tcolorbox}
\usepackage{graphicx}
\usepackage{subcaption}

\newcolumntype{L}[1]{>{\raggedright\let\newline\\\arraybackslash\hspace{0pt}}m{#1}}
\newcolumntype{C}[1]{>{\centering\let\newline\\\arraybackslash\hspace{0pt}}m{#1}}
\newcolumntype{R}[1]{>{\raggedleft\let\newline\\\arraybackslash\hspace{0pt}}m{#1}}

\begin{document}

\newtcolorbox{findingbox}{
colback=gray!6,
colframe=black!55,
boxrule=0.4pt,
arc=2pt,
left=5pt,
right=5pt,
top=4pt,
bottom=4pt,
fontupper=\small
}

\begin{frontmatter}


\title{Investigating Developer-Reported Software Security Testing Challenges}

\author[ua]{Md Erfan}
\ead{merfan@crimson.ua.edu}

\author[ua]{Ahmed Ryan}
\ead{aryan9@crimson.ua.edu}

\author[ua]{Md Rayhanur Rahman\corref{cor1}}
\ead{mrahman87@ua.edu}

\cortext[cor1]{Corresponding author}

\affiliation[ua]{
  organization={Department of Computer Science, University of Alabama},
  city={Tuscaloosa},
  state={AL},
  country={USA}
}

\begin{abstract}
\textbf{Context:} Software security testing (SST) is essential for identifying vulnerabilities and improving software security. However, developers often face challenges when selecting tools, configuring test environments, interpreting security scanner outputs, testing authentication workflows, and acting on reported vulnerabilities in real-world software development practice.

\noindent \textbf{Objectives:} This study empirically characterizes developer-reported SST challenges in Stack Overflow (SO) discussions and provides a structured understanding of their prevalence, difficulty, temporal evolution, co-occurrence, and taxonomy stability across developer support contexts.

\noindent \textbf{Methods:} We conduct a large-scale empirical study of 17,743 SO questions collected using SST-related keywords. Through manual labeling, we identify 582 SST-related questions and develop a taxonomy consisting of 8 categories and 31 subcategories. We then analyze the identified challenges using prevalence, difficulty, correlation, temporal trend, co-occurrence, and held-out stability analyses, as well as broader developer-facing security challenge patterns.

\noindent \textbf{Results:} Developers discuss security finding interpretation and reliability, security testing guidance and tool selection, authentication and authorization testing, and security tool integration and automation. Validation-related challenges often require more technical context and longer resolution time. Security finding interpretation and remediation-related actionability show increasing trends over time. Co-occurrence analysis shows that false positives frequently appear with explainability, while integration challenges often appear with tool suggestions and documentation/resources.

\noindent \textbf{Conclusion:} Developer-reported SST challenges extend beyond vulnerability detection and include workflow, configuration, interpretation, and remediation concerns. These findings can help researchers, practitioners, educators, and tool providers better understand developer needs and improve SST tool usability, documentation, result interpretation, and remediation support.
\end{abstract}


\end{frontmatter}

\section{Introduction}
Security testing is an essential practice for assessing whether software can resist adversarial inputs, misuse cases, and violations of security assumptions~\cite{d2015correctness,khamaiseh2017software}. Unlike functional testing, which focuses on expected behavior under normal conditions, security testing examines software behavior under malicious, unexpected, or misuse-oriented conditions. Modern applications now process sensitive data, depend on third-party libraries, expose APIs, and run in environments where attackers can exploit weak inputs, misconfigurations, vulnerable dependencies, or broken authentication logic~\cite{schott2025bytecode,chenpatch}. Therefore, \gls{sst} has become an essential complement to functional testing and an important part of the \gls{sdlc}. Developers commonly use automated \gls{sst} techniques such as \gls{sast}~\cite{li2020vulnerabilities}, \gls{dast}~\cite{sharma2021review,dencheva2022comparative}, fuzz testing~\cite{zhu2022fuzzing,li2018fuzzing}, and dependency scanning to identify security risks during development. However, applying \gls{sst} in practice remains challenging because developers must select appropriate tools, configure them correctly, interpret their outputs, and decide how to act on reported findings~\cite{shen2026llm,du2026reducing,ameen2026qasecclaw, morshed2026use}. As a result, developers often turn to question-and-answer platforms such as \gls{sta} to seek help and discuss practical challenges encountered during \gls{sst} in modern software development workflows today.

Prior work has examined \gls{sst} techniques and security testing tools, including \gls{sast}, \gls{dast}, \gls{iast}, fuzz testing, penetration testing, and security checks across the \gls{sdlc}~\cite{sanne2024investigations,dowd2006art,tondel2008learning,brucker2017enable,soares2023perceptions,rahman2025sok}. Empirical studies also show that security testing tools can be difficult to adopt in practice because developers must manage configuration issues, workflow integration, false positives, and result interpretation~\cite{marcilio2019automatically,marcilio2019static,rangnau2020continuous,freitas2024gash,zheng2025github}. However, existing work provides limited empirical evidence on how developers themselves describe \gls{sst} challenges when they seek help in community-based settings such as \gls{sta}. This gap matters because developer discussions show practical challenges that may not be visible from tool evaluations or repository mining alone. Understanding these challenges can help researchers identify practice-driven research problems, help tool providers improve usability and reporting, and help practitioners find clearer guidance for applying \gls{sst} in real development workflows across diverse technical contexts.

\textit{
The goal of this article is to provide empirical evidence on developer-reported \gls{sst} challenges on \gls{sta} and explain how these challenges affect real development workflows, helping researchers define practice-driven challenges, practitioners recognize recurring challenges, and tool providers improve documentation, reporting, and workflow support.
}

We achieve the goal by addressing the following research questions (RQs).
\begin{itemize}[leftmargin=*]

    \item \textbf{RQ1: (SST Taxonomy)} What challenges do developers face in \gls{sst}?
    \item \textbf{RQ2: (Problem Characteristics)} What are the prevalent characteristics and trends of the identified \gls{sst} challenges? We break down \textbf{RQ2} into five sub-research questions:
    \begin{itemize}
        \item \textbf{RQ2.1 (Prevalence):} Which challenges are the most prevalent among developers across community-based \gls{sst} discussions?
        
        \item \textbf{RQ2.2 (Difficulty):} Which \gls{sst} challenges are the most difficult in terms of obtaining satisfactory resolutions?
        
        \item \textbf{RQ2.3 (Correlation):} What is the relationship between the prevalence of \gls{sst} challenges and their perceived technical difficulty?
        
        \item \textbf{RQ2.4 (Temporal Trends):} How do \gls{sst} challenges trend and evolve over time?
        
        \item \textbf{RQ2.5 (Co-occurrence):} Which \gls{sst} challenge subcategories co-occur in questions?
    \end{itemize}
    \item \textbf{RQ3: (Taxonomy Stability)} To what extent does the proposed \gls{sst} taxonomy remain stable when applied to temporally unseen developer questions?
\end{itemize}

We address these RQs through a mixed qualitative and quantitative analysis of \gls{sta} discussions. We first collected 17,743 questions using \gls{sst}-related keywords derived from academic literature and practitioner-oriented sources. We then applied inclusion and exclusion criteria through a manual labeling process and identified 582 questions focused on \gls{sst} challenges. For RQ1, we used inductive coding and iterative category refinement to construct a practitioner-derived taxonomy of developer challenges. For RQ2, we measured prevalence and difficulty using community engagement and resolution metrics, analyzed metric relationships with Pearson and Spearman correlations, examined temporal trends with the Cox-Stuart test, and studied co-occurring challenges. For RQ3, we applied the finalized taxonomy to temporally unseen questions and measured whether these questions could be mapped to existing categories and subcategories without requiring taxonomy changes.

\noindent We list our contributions as follows:
\begin{itemize}[leftmargin=*]

\item A practitioner-derived taxonomy of developer-reported \gls{sst} challenges on \gls{sta}, consisting of 8 categories and 31 subcategories.

\item An empirical characterization of \gls{sst} challenges using prevalence, difficulty, correlation, temporal trend, and co-occurrence analyses based on \gls{sta} engagement and resolution metrics.

\item A held-out stability evaluation showing that temporally unseen \gls{sta} questions can be mapped to the proposed taxonomy without requiring new categories or subcategories.

\item A set of implications for researchers, practitioners, educators, and tool providers based on the observed \gls{sst} challenge patterns.

\end{itemize}

\noindent We publicly release our source code, datasets, and experimental artifacts on Figshare at \\ \textcolor{red}{https://figshare.com/s/17e56c96905c8ab423b4}.

The remainder of the paper is organized as follows. Section~\ref{sec:key_concepts} defines the key concepts used in this study. Section~\ref{sec:related_work} reviews related work, and Section~\ref{sec:approach} describes our research approach. Sections~\ref{sec:rq1_taxonomy}, \ref{sec:rq2_popularity_difficulty}, and \ref{sec:rq3_taxonomy_stability} present the findings for RQ1, RQ2, and RQ3, respectively. Section~\ref{sec:discussions} discusses implications and future directions. Section~\ref{sec:threat_to_validity} presents threats to validity, and Section~\ref{sec:conclusion} concludes the paper.

\section{Key Concepts}
\label{sec:key_concepts}

This section defines the key concepts used throughout the study. These concepts provide the methodology for constructing the taxonomy and analyzing developer-reported \gls{sst} challenges.

\textbf{Software Security Testing.}
\gls{sst} refers to the systematic assessment of software systems to identify, validate, and evaluate security weaknesses that may expose the system to unauthorized access, data leakage, privilege escalation, denial of service, or other security violations. Unlike general functional testing, which focuses on expected behavior, \gls{sst} focuses on security-relevant behavior under adversarial inputs, misuse cases, vulnerable configurations, and threat conditions~\cite{bui2025systematic, nawaz2025cluster}. Common \gls{sst} activities include static analysis, dynamic analysis, fuzzing, penetration testing, dependency scanning, authentication testing, and security checks in CI/CD workflows across modern development and deployment environments~\cite{sanne2024investigations,dowd2006art,tondel2008learning,brucker2017enable,soares2023perceptions,rahman2025sok}.

\textbf{Qualitative coding and thematic analysis.}
Qualitative coding is a systematic qualitative analysis process in which descriptive labels, or codes, are assigned to textual data to identify concepts, patterns, and meanings grounded in the data~\cite{corbin2014basics,cruzes2011recommended}. \textbf{Inductive coding} is a coding approach in which codes are derived from the data rather than from a predefined classification scheme~\cite{treude2024qualitative, chen2024computational}. The coding is commonly used when the objective is to discover recurring concepts from textual evidence. \textbf{Thematic analysis} is a qualitative method for identifying, analyzing, and organizing patterns of meaning across a dataset~\cite{braun2006using,braun2019reflecting}. In thematic analysis, related codes are interpreted and grouped into broader themes that represent higher-level concepts. \textbf{Reflexive thematic analysis} emphasizes researcher interpretation, iterative refinement, and discussion during the development of themes~\cite{byrne2022worked, braun2021one}. Together, qualitative coding and thematic analysis provide a structured way to move from detailed textual observations to broader conceptual categories.

\textbf{Stack Overflow.}
\gls{sta} is a developer-centered Q\&A platform within the Stack Exchange network where programmers ask technical questions, share solutions, and discuss practical software development problems. The platform provides question titles and bodies, answers, comments, scores, views, tags, creation dates, and accepted-answer information. These data make \gls{sta} useful for studying how developers describe problems, seek help, and receive community responses across different technical domains~\cite{stackexchange_data,stackoverflow_site,rahman2018questions,tahir2020large,zhang2024developers}.

\section{Related Work}
\label{sec:related_work}

This section discusses related work on \gls{sst} techniques, security testing tools, developer challenges, community-based software engineering discussions, and taxonomy-based empirical analysis.

\subsection{Security Testing Techniques and Tooling Foundations}

Foundational research defines the technical landscape of \gls{sst} and explains how security testing techniques are used across the \gls{sdlc}. Prior surveys and systematic reviews summarize major approaches such as \gls{sast}, \gls{dast}, \gls{iast}, fuzzing, penetration testing, vulnerability assessment, and security checks in modern development workflows~\cite{sanne2024investigations,dowd2006art,tondel2008learning,brucker2017enable,soares2023perceptions,rahman2025sok,deshmukh2024automated,paule2018securing,bhardwaj2024securing}. Related taxonomic studies also classify security assessment tools and computer security incidents~\cite{shostack1999towards,kiltz2007taxonomy}. These studies establish important foundations for understanding tool capabilities, vulnerability targets, coverage, automation, and reporting trade-offs. However, they provide limited insight into how developers describe practical \gls{sst} challenges when configuring tools, interpreting findings, testing protected functionality, or deciding whether reported vulnerabilities are actionable. Our RQ1 builds on this foundation by constructing a taxonomy of developer-reported \gls{sst} challenges.

\subsection{Empirical Studies of Security Testing Tools and Workflows}

Empirical studies examine how security testing tools operate in practice. Prior work analyzes \gls{sast} tools, vulnerability scanners, DevSecOps workflows, GitHub Actions, and CI/CD failures, showing that practical adoption is affected by false positives, configuration complexity, workflow failures, warning interpretation, and integration barriers~\cite{marcilio2019automatically,marcilio2019static,rangnau2020continuous,freitas2024gash,zheng2025github,ananya2025ai}. These studies provide valuable evidence about tool behavior and workflow reliability, but they usually focus on specific tools, repositories, or workflow settings. Prior work therefore provides limited insight into how developers report and experience \gls{sst} challenges across tools, testing contexts, authentication scenarios, dependency scanning, reporting, and remediation. Our RQ2 extends this line of work by measuring the prevalence, difficulty, correlation, temporal trends, and co-occurrence patterns of \gls{sst} challenges in community-based developer support environments.

\subsection{Developer Challenges in Community Q\&A Platforms}

Community Q\&A platforms such as \gls{sta} provide a rich source of practitioner-reported problems. Prior studies have used \gls{sta} and related developer platforms to analyze challenges in domains such as checked-in secrets, infrastructure-as-code, code smells, and software maintenance~\cite{basak2023challenges,rahman2018questions,tahir2020large,zhang2024developers}. These studies show that developer questions can reveal recurring challenges, missing documentation, tool misunderstandings, configuration issues, and knowledge gaps that may not be visible from repository mining alone. They also demonstrate the value of manual labeling and taxonomy construction for understanding developer needs. However, existing community-based studies mainly focus on adjacent software engineering or security topics rather than \gls{sst} as a dedicated domain. Our study addresses this gap by analyzing \gls{sta} discussions specifically related to \gls{sst} challenges.

\subsection{Taxonomy Construction, Trend Analysis, and Stability Validation}

Taxonomy-based empirical studies organize developer challenges into interpretable categories and subcategories. Prior work has used manual coding, iterative category refinement, inter-rater agreement, prevalence analysis, difficulty analysis, temporal trends, and validation to characterize developer problems from discussions, issues, and repositories~\cite{rahman2018questions,tahir2020large,zhang2024developers,basak2023challenges}. Although these methods have been applied in related software engineering domains, \gls{sst} has received limited attention as a target for taxonomy construction, problem characterization, and stability validation. Existing studies provide foundations for individual security testing techniques, tool evaluations, or adjacent developer challenges, but they do not systematically characterize \gls{sst} challenges across taxonomy, prevalence, difficulty, correlation, temporal trends, co-occurrence, and held-out stability. Our RQ3 addresses this issue by evaluating whether the proposed \gls{sst} taxonomy remains stable when applied to temporally unseen developer questions.

Overall, prior work provides important foundations for understanding \gls{sst} techniques, tool behavior, workflow barriers, and taxonomy-based analysis; however, developer-reported \gls{sst} challenges on \gls{sta} remain underexplored as a dedicated empirical research focus.

\section{Approach}
\label{sec:approach} 

Our methodology follows a multi-stage empirical process. We first identify \gls{sst}-related keywords from academic papers and practitioner resources. We then use these keywords to collect \gls{sta} questions and apply criteria to construct the final dataset. Next, we analyze the filtered dataset to construct the taxonomy, measure problem characteristics, and evaluate taxonomy stability. Finally, we report the resulting \gls{sst} challenge taxonomy and the main empirical findings.

\subsection{Data Collection}
This subsection describes our process for collecting and curating \gls{sta} questions by selecting \gls{sta} as the data source, constructing an \gls{sst}-related keyword corpus, and applying inclusion and exclusion criteria to obtain the final study dataset.

\subsubsection{Q\&A Site Selection}

We select \gls{sta} as the primary data source because it provides large-scale developer discussions with question text, answers, comments, scores, views, tags, creation dates, and accepted-answer information. These data allow us to analyze both the content of developer-reported \gls{sst} challenges and their resolution signals. At the time of our analysis, \gls{sta} contained approximately 24 million questions, 36 million answers, 30 million users, and a 70\% answer rate, making it suitable for studying practical \gls{sst}-related challenges across diverse software development contexts.

\subsubsection{Keyword Identification and Corpus Construction}

At the time of our analysis, \gls{sta} contained approximately 24 million questions covering diverse software development topics. We constructed a keyword corpus to retrieve questions related to \gls{sst}. The keyword corpus draws from two complementary sources, academic literature and grey literature, so that the search terms reflect both research terminology and practitioner vocabulary. We followed the steps below to improve search coverage and relevance.

\begin{enumerate}[(a)]

    \item \textbf{Literature-driven keyword identification ($\mathcal{C}_1$):}
    We first constructed an academic keyword corpus from security testing research published in high-impact software engineering and security venues. We selected venues based on their relevance to software engineering, testing, and systems security, named ICSE~\cite{DBLP:conf/icse/2024}, ISSTA~\cite{DBLP:conf/issta/2024}, IEEE S\&P~\cite{DBLP:conf/sp/2024}, and USENIX Security~\cite{DBLP:conf/uss/2024}. We then searched the digital archives of these venues using root terms related to \gls{sst}: \textit{security testing}, \textit{vulnerability}, \textit{vulnerabilities}, \textit{static analysis}, \textit{dynamic analysis}, \textit{penetration testing}, \textit{fuzz testing}, and \textit{fuzzing}. For each relevant article, we extracted the author-defined keywords when available. When author-defined keywords were not available, we reviewed the abstract and extracted representative terms related to the study topic, technique, tool, or testing context, following a similar keyword-extraction practice used in prior software engineering mapping research~\cite{santhanam2022bots, petersen2015guidelines}. This process produced the literature-driven keyword corpus ($\mathcal{K}_1$), which captures terminology commonly used in academic \gls{sst} research.

    \item \textbf{Grey-literature-driven keyword identification and validation ($\mathcal{C}_2$):}
    We next incorporated grey literature to capture practitioner terminology that may not appear frequently in academic publications. Grey literature refers to technical content produced outside traditional academic publishing, such as industry reports, security blogs, technical documentation, and vulnerability disclosures~\cite{tian2026systematizing}. We used LLM-assisted discovery through Gemini~\cite{google_gemini_2025}, ChatGPT~\cite{openai_chatgpt_2025}, and Perplexity~\cite{perplexity_ai_2025} with the prompt \texttt{"top software security testing blogs and technical reports"} to identify candidate practitioner-oriented sources. From the retrieved sources, we manually selected materials that focused on \gls{sst}, including security blogs such as \textit{Google Project Zero}~\cite{google_project_zero} and \textit{PortSwigger Research}~\cite{portswigger_research}, and industry reports such as OWASP Top 10~\cite{owasp_top_10} and the Verizon DBIR~\cite{verizon_dbir_2024}. The first and third authors discussed the selected sources and retained only sources that were directly relevant to security testing practices. We then extracted \gls{sst}-related terms from these sources to form the grey-literature-driven keyword corpus ($\mathcal{K}_2$). The practitioner-oriented corpus complements the academic corpus by incorporating terminology used in practitioner communities and industry-facing security resources.
\end{enumerate}

\subsubsection{Final Keyword Compilation and Validation}

After constructing the literature-driven corpus ($\mathcal{K}_1$) and the grey-literature-driven corpus ($\mathcal{K}_2$), we combined both corpora to produce the final keyword list. This step ensured that the search terms captured both academic terminology and practitioner vocabulary related to \gls{sst}. The final keyword compilation and validation process involved the following steps.

\begin{enumerate}[(a)]

    \item \textbf{Keyword normalization, aggregation, and validation.}
    We normalized the collected keywords to ensure consistency during search. The same \gls{sst} concept may appear under different terms, such as \textit{DAST} and \textit{Dynamic Application Security Testing}; therefore, we grouped synonyms and equivalent expressions to reduce redundancy while preserving coverage. We also removed overly generic terms, such as \textit{software} and \textit{testing}, because these terms could retrieve broad software engineering discussions rather than security-specific testing challenges. After normalization, the first and third authors independently reviewed the aggregated keyword list and assessed whether each keyword was related to developer-facing \gls{sst} challenges. This review helped distinguish keywords that reflected actionable security testing concerns from terms likely to retrieve irrelevant discussions, such as abstract security theory, general installation issues, or broad tool usage questions. The first and third authors resolved disagreements through discussion and selected the final retained keywords by consensus.
    \item \textbf{Final keyword list ($\mu$).} After normalization and manual validation, we combined the retained keywords from $\mathcal{C}_1$ and $\mathcal{C}_2$ to form the final keyword list ($\mu$). The final keyword list covers major \gls{sst} techniques, tools, and practices: \textit{Static Application Security Testing (SAST)}, \textit{Dynamic Application Security Testing (DAST)}, \textit{Interactive Application Security Testing (IAST)}, \textit{Runtime Application Security Testing (RAST)}, \textit{Fuzz Testing}, \textit{Vulnerability Scanning}, \textit{Penetration Testing}, \textit{Security Code Review}, \textit{Threat Modeling}, \textit{Security in CI/CD}, and \textit{OWASP}.

\end{enumerate}

\subsubsection{Dataset Curation}

After constructing the final keyword list ($\mu$), we curated the dataset through three steps: dataset construction, data definition, and inclusion/exclusion classification.

\begin{enumerate}[(a)]

\item \textbf{Dataset construction.}

We retrieved \gls{sst}-related questions from \gls{sta} using the final keyword list ($\mu$). Although \gls{sta} data can be accessed through the REST API, SQL-based queries, and third-party scrapers, we used the Stack Exchange Data Explorer~\cite{stackexchange_data} because the platform provides reproducible SQL-based access to the \gls{sta} data schema and supports large-scale question retrieval. The REST API was not used because rate limits and daily quotas would make large-scale retrieval inefficient~\cite{stackexchange_api_throttle}. We queried questions posted between January, 2015, and August, 2025. For each keyword in $\mu$, we searched the \textit{Posts} table and retained records with \textit{PostTypeId = 1}, which corresponds to questions. The results from all keyword searches were combined to form the initial dataset $D$. Duplicate questions retrieved by multiple keywords were removed using the \gls{sta} post identifier before constructing the final analysis dataset.

\item \textbf{Data definition.}
Each question in $D$ was treated as one data record. For each record, we collected the Post ID, Title, and Body. The Post ID served as a unique identifier for retrieving metadata such as view count, score, answer count, and accepted-answer status. The Title and Body were used together during manual analysis because they jointly describe the technical context, attempted solution, observed behavior, and developer-facing \gls{sst} challenge.

\item \textbf{Inclusion and exclusion criteria.}
We developed inclusion and exclusion criteria to filter questions related to \gls{sst} challenges. The criteria were designed to distinguish developer-facing \gls{sst} challenges from general security questions, routine testing issues, implementation-only questions, and general tool-configuration problems without clear \gls{sst} intent.

From the initial dataset $D$, we randomly selected 370 questions, corresponding to approximately 2\% of the candidate questions, to develop the inclusion and exclusion criteria. This sample was used to iteratively identify recurring inclusion and exclusion decisions until the criteria reached thematic saturation. The final criteria include three inclusion criteria (IC1--IC3) and three exclusion criteria (EC1--EC3), as shown in Table~\ref{tab:inclusion_exclusion_selection_criteria}. The inclusion criteria capture questions about \gls{sst} tools, platforms, standards, resources, vulnerability-detection challenges, and CI/CD-based security testing workflows. The exclusion criteria remove questions about general software development, access-control implementation without testing intent, and general troubleshooting or configuration problems without clear \gls{sst} relevance.

\begin{table}[!t]
\centering
\caption{Inclusion and Exclusion Criteria}
\label{tab:inclusion_exclusion_selection_criteria}
\small
\setlength{\tabcolsep}{4pt}
\renewcommand{\arraystretch}{1.12}
\begin{tabularx}{\textwidth}{
>{\centering\arraybackslash}p{0.08\textwidth}
>{\raggedright\arraybackslash}p{0.25\textwidth}
>{\arraybackslash}X
}
\toprule
\textbf{ID} & \textbf{Criterion Title} & \textbf{Criterion Description} \\
\midrule

\multicolumn{3}{l}{\textbf{Inclusion Criteria}} \\
\midrule

\textbf{IC1} &
\textbf{Challenges in SST-related Tooling, Platforms, Standards, and Resources} &
The question must involve technical or logistical challenges in \gls{sst}, including tool scalability, licensing, vendor support, platform compatibility, resources, integration into development/deployment environments, lack of standardized \gls{sst} practices, or unclear vulnerability reports, false positives, and insufficient output details. \\

\textbf{IC2} &
\textbf{Challenges in Detecting Security Vulnerabilities via SST} &
The question focuses on difficulties in detecting security vulnerabilities, such as SQL injection, XSS, buffer overflows, or Man-in-the-Middle attacks, during the execution of a security software test in practical developer testing scenarios. \\

\textbf{IC3} &
\textbf{Challenges in Integrating SST into Automated Development Workflows} &
The question addresses challenges in setting up, configuring, or running security tests within automated workflows, such as CI/CD pipelines and other continuous delivery environments. \\

\midrule
\multicolumn{3}{l}{\textbf{Exclusion Criteria}} \\
\midrule

\textbf{EC1} &
\textbf{General Software Development and Implementation without SST Focus} &
Exclude routine implementation, development, environment, GUI, browser compatibility, integration, syntax, file-path, dependency-installation, remediation, patching, or vulnerability-prevention questions without clear \gls{sst} intent. \\

\textbf{EC2} &
\textbf{Access Control and Session Management without SST Focus} &
Exclude questions focused only on login systems, authentication, token handling, session tracking, or access-control implementation when there is no clear emphasis on testing or security assessment. \\

\textbf{EC3} &
\textbf{General Troubleshooting or Tool Configuration without SST Focus} &
Exclude general maintenance, configuration, dependency, library, plugin, crash, connection, compliance, or containerized deployment issues that lack a clear security-testing focus. \\

\bottomrule
\end{tabularx}
\end{table}

\item \textbf{Inclusion and exclusion classification.}
We performed a two-round labeling process to identify questions that satisfied the inclusion criteria and did not meet the exclusion criteria. Following established qualitative coding practices~\cite{kou2022sosum}, the first and third authors first used the 370-question pilot sample to refine the criteria and resolve ambiguous cases. For the main classification, the same two authors independently reviewed 10\% (1,738) randomly selected questions using the criteria in Table~\ref{tab:inclusion_exclusion_selection_criteria}. The initial round produced a Cohen's kappa coefficient of $\kappa = 0.61$, indicating moderate agreement. The annotators then met to discuss disagreements and formalize ambiguous cases into a more explicit labeling guide. Using the refined guide, the annotators re-evaluated the conflicting cases, increasing the agreement to $\kappa = 0.69$. After resolving the remaining disagreements through discussion, we applied the finalized criteria to the remaining candidate questions. 

\end{enumerate}

\subsection{RQ1: Construction of the SST Challenge Taxonomy}

We constructed the taxonomy from the 582 filtered questions using inductive coding and reflexive thematic analysis. The first and third authors independently reviewed question titles and bodies, assigned open codes to capture developer-facing \gls{sst} challenges, and treated these codes as candidate subcategories. They then compared codes, merged overlapping labels, refined unclear terms, and grouped similar challenges. For example, \textit{tool capability} was refined as \textit{features}, while \textit{best practices}, \textit{secure coding guidance}, and \textit{tool applicability} were grouped under \textit{guidance}. The initial raw agreement was 83\%, and all disagreements were resolved through discussion, resulting in 100\% final agreement.

Next, the authors refined the subcategories and grouped them into broader categories based on conceptual similarity. The second author reviewed the proposed structure, and all authors discussed and finalized the taxonomy through consensus. The final taxonomy organizes developer-reported \gls{sst} challenges into categories and subcategories, where each category represents a broad area of \gls{sst} practice and each subcategory captures a specific developer-facing challenge. We maintain traceability by assigning each \gls{sta} question an incremental identifier from Q1 to Q582.

\subsection{RQ2: Problem Characteristics}

We characterize \gls{sst} challenges using prevalence, difficulty, correlation, temporal trend, and co-occurrence analyses at the category and subcategory levels.

\subsubsection{Prevalence Metrics (RQ2.1). Which \gls{sst} challenges are the most prevalent among the developer community?}
\label{ref:popularity_metrics}

We measure the prevalence and visibility of \gls{sst} challenges using four \gls{sta} engagement metrics commonly used in prior empirical studies: views, scores, answers, and comments~\cite{rahman2018questions,tahir2020large,zhang2024developers}. \textbf{avgView} captures the average visibility of questions in a subcategory, \textbf{avgScore} captures their perceived community value, \textbf{avgAnswer} captures community response, and \textbf{avgComment} captures clarification and discussion before formal answer-level resolution.

\subsubsection{Difficulty Metrics (RQ2.2). Which \gls{sst} challenge categories exhibit more difficulty in terms of obtaining satisfactory resolutions?}
\label{ref:difficulty_metrics}

We measure the resolution difficulty of \gls{sst} challenges using five complementary \gls{sta} metrics: answer availability, accepted-answer status, response time, accepted-answer time, and question length. \textbf{AnswerRate} captures whether questions in a subcategory receive any answer, \textbf{AcceptRate} captures satisfactory resolution, \textbf{TimeFirstAnswer} captures the speed of initial community response, \textbf{TimeAcceptedAnswer} captures the time required to obtain an accepted solution, and \textbf{TextSize} captures the amount of technical context provided in the question body, such as logs, configurations, stack traces, tool outputs, or reproduction steps.

\subsubsection{Correlation (RQ2.3). What is the relationship between the prevalence of a challenge and its perceived technical difficulty?}

We analyze the relationship between prevalence and difficulty by first applying Min-Max normalization to place all metrics on a common 0--1 scale~\cite{juszczak2002feature}. We then compute Pearson correlation~\cite{benesty2009pearson} and Spearman's rank correlation~\cite{sedgwick2014spearman} to examine both linear and rank-based relationships between prevalence and difficulty metrics. This analysis shows whether highly visible or frequently discussed \gls{sst} challenges are also easier or harder to resolve.

\subsubsection{Temporal Trend (RQ2.4). How do \gls{sst} challenge categories trend and evolve over time?}

We analyze temporal evolution at the category level by computing the normalized monthly proportion of questions in each \gls{sst} category, as shown in Equation~\ref{eq:temporal_trend}. This normalization accounts for monthly variation in the total number of \gls{sst}-related questions.

\begin{equation}
TT(x, m) = 
\frac{\textit{count of category } x \text{ in month } m}
{\textit{total \gls{sst} questions in month } m}
\label{eq:temporal_trend}
\end{equation}

Here, $TT(x,m)$ denotes the monthly proportion of questions for category $x$ in month $m$. We then apply the Cox-Stuart test~\cite{cox1955some} at a 95\% confidence level ($p < 0.05$) to classify each category as increasing, decreasing, or consistent over time.

\subsubsection{Co-occurrence Analysis (RQ2.5). Which \gls{sst} challenge subcategories co-occur in developer questions?}

We analyze co-occurrence because a single \gls{sta} question may involve multiple \gls{sst} challenges, such as integration with configuration or false positives with explainability. We model each question as a transaction containing one or more assigned subcategories and treat two subcategories as co-occurring when they appear in the same question. For each pair of subcategories $a$ and $b$, we compute raw co-occurrence count, Jaccard similarity~\cite{jaccard1901etude,jackson1989similarity}, and lift~\cite{agrawal1993mining,brin1997beyond}, as shown in Equations~\ref{eq:cooccurrence_count}--\ref{eq:lift}.

\begin{equation}
CoOccur(a,b) = |Q_a \cap Q_b|
\label{eq:cooccurrence_count}
\end{equation}

\begin{equation}
Jaccard(a,b) = \frac{|Q_a \cap Q_b|}{|Q_a \cup Q_b|}
\label{eq:jaccard}
\end{equation}

\begin{equation}
Lift(a,b) = \frac{P(a,b)}{P(a)P(b)}
= \frac{|Q_a \cap Q_b|/N}{(|Q_a|/N)(|Q_b|/N)}
\label{eq:lift}
\end{equation}

Here, $Q_a$ and $Q_b$ denote the sets of questions assigned to subcategories $a$ and $b$, and $N$ denotes the total number of questions. Raw count captures frequent pairs, Jaccard captures relative overlap, and lift captures whether two subcategories appear together more often than expected by chance. We visualize these relationships using category-level and subcategory-level heatmaps, where darker cells indicate stronger co-occurrence.

\subsection{RQ3: Taxonomy Stability on Temporally Unseen Questions}
\label{subsec:taxonomy_stability}

We evaluate taxonomy stability by testing whether the proposed \gls{sst} taxonomy can classify temporally unseen \gls{sta} questions without requiring new categories or subcategories. The taxonomy was constructed using \gls{sst}-related questions collected up to August 2025. We then collected a separate held-out set of \gls{sst}-related questions posted after August 2025. These held-out questions were excluded from keyword refinement, inclusion/exclusion filtering, open coding, subcategory refinement, and category construction, making them an independent validation sample.

We manually map each held-out question to the existing taxonomy. A question is considered covered if its main developer challenge can be assigned to at least one existing subcategory. Because some questions involve multiple concerns, we allow multiple subcategory assignments when appropriate. The main stability criterion is whether the taxonomy explains each held-out question without requiring a new category or subcategory. We measure held-out coverage using Equation~\ref{eq:heldout_coverage}, where $H$ denotes the held-out question set and $Covered(H)$ denotes the subset of held-out questions mapped to at least one existing taxonomy subcategory.

\begin{equation}
Coverage(H) = \frac{|Covered(H)|}{|H|} \times 100
\label{eq:heldout_coverage}
\end{equation}

\section{Findings from RQ1: Taxonomy of SST Challenges}
\label{sec:rq1_taxonomy}

\begin{figure*}[!t] 
    \centering
    \includegraphics[width=0.98\textwidth]{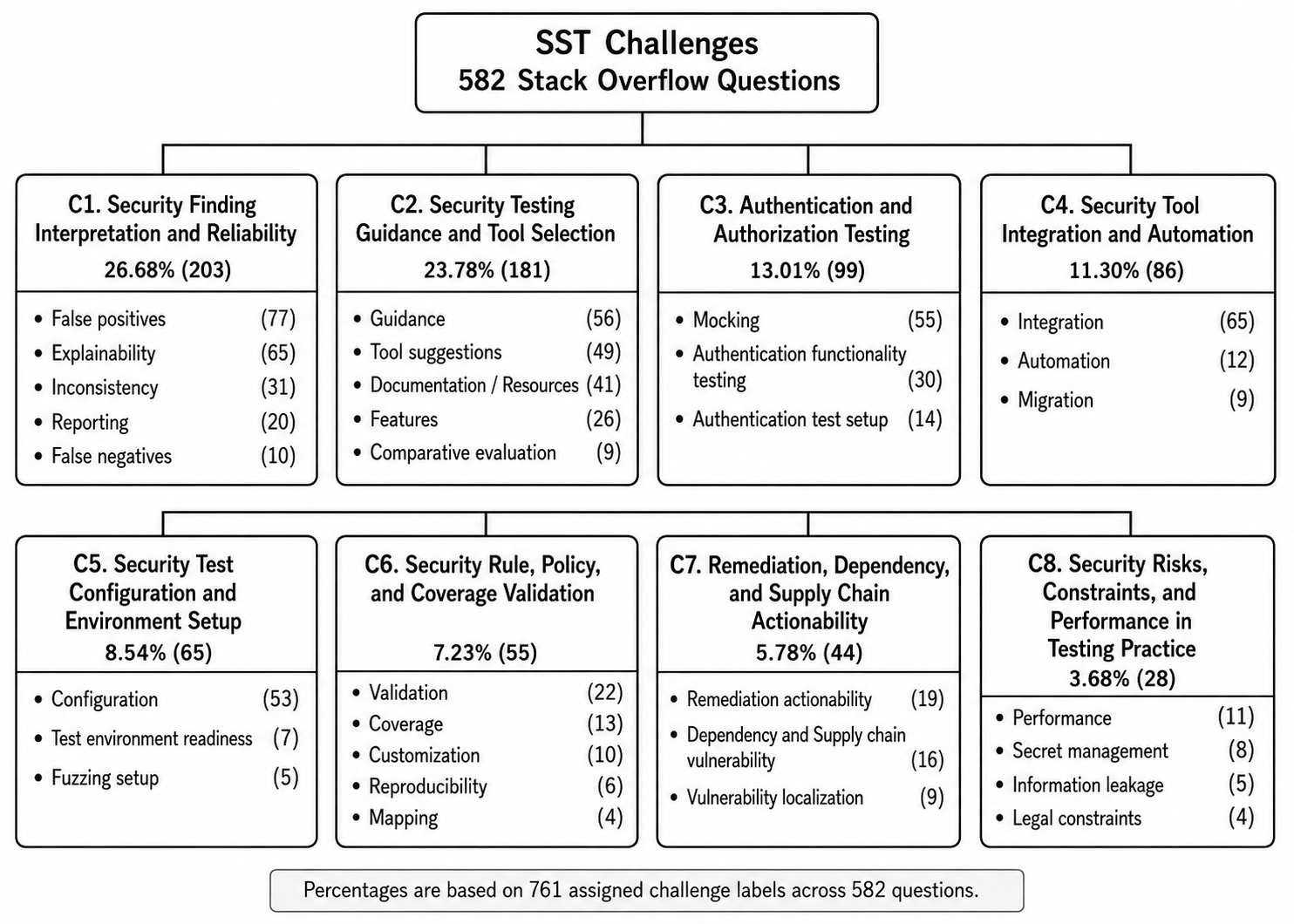} 
    \caption{Hierarchical taxonomy of developer-reported \gls{sst} challenges.}
    \label{fig:sst-taxonomy}
\end{figure*}

Figure~\ref{fig:sst-taxonomy} presents the hierarchical taxonomy of \gls{sst} challenges, consisting of 8 categories and 31 subcategories derived from 582 \gls{sta} questions. The taxonomy shows that developer-reported \gls{sst} challenges span the broader testing workflow, including tool selection, configuration, authentication testing, finding interpretation, integration, dependency analysis, and remediation.

\begin{findingbox}
\textbf{Finding 1}. \textit{Developers discuss a wide range of \gls{sst} challenges, which can be grouped into a hierarchical taxonomy of 8 categories and 31 subcategories. Security finding interpretation and reliability is discussed the most by developers (26.68\%), followed by security testing guidance and tool selection (23.78\%).}
\end{findingbox}

We next discuss each category with its subcategories and representative examples to explain the taxonomy in detail and clarify its practical meaning.

\subsection{C1: Security Finding Interpretation and Reliability (26.68\%)}

C1 captures challenges developers face when \gls{sst} tools produce findings that are difficult to understand, verify, report, reproduce consistently, or trust. Scanner findings often require developer judgment before they can be treated as actionable issues. Developers must determine whether a warning is explainable, whether a reported issue is a false positive, whether the report contains enough information, whether results are consistent across environments, and whether expected vulnerabilities were missed. As shown in Table~\ref{tab:finding-interpretation-reliability-subtheme}, C1 includes five subcategories: false positives, explainability, inconsistency, reporting, and false negatives.

\begin{table}[t]
\centering
\caption{C1 - Security Finding Interpretation and Reliability}
\label{tab:finding-interpretation-reliability-subtheme}
\small
\setlength{\tabcolsep}{4pt}
\renewcommand{\arraystretch}{1.15}
\begin{tabularx}{\textwidth}{
>{\raggedright\arraybackslash}p{0.21\textwidth}
Y
}
\hline
\textbf{Name} & \textbf{Explanation} \\
\hline

\textbf{False positives (77)} &
Developers face false-positive challenges when \gls{sst} tools report findings that may not represent real or exploitable security issues. Developers must decide whether a reported finding is a real vulnerability or a tool-generated false alarm, such as Fortify reporting a JavaScript issue around \texttt{window.opener.location.href} during review of browser-based vulnerability reports in practice \textbf{(Q142~\cite{Q142}: ``window.opener.location.href | Fortify scan false positivee'')}.\\
\hline

\textbf{Explainability (65)} &
Developers face explainability challenges when \gls{sst} tools produce findings, reports, warnings, or vulnerability details that are difficult to interpret or trace. These challenges appear when developers cannot understand low-level diagnostic outputs, such as unresolved symbols in an AddressSanitizer stack trace \textbf{(Q4~\cite{Q4}: ``Unresolved symbol in stacktrace when using GCC 4.8.2's Asan'')}.\\
\hline

\textbf{Inconsistency (31)} &
Developers face inconsistency challenges when \gls{sst} tools produce different, missing, or unexpected results across interfaces, environments, scanners, pipelines, or repeated runs. These challenges appear when a CI/CD security scan does not report the same vulnerabilities shown by the corresponding web-based scanner under equivalent dependency and configuration settings \textbf{(Q184~\cite{Q184}: ``Jenkins Snyk Task Not Finding Vulnerabilities Found via Snyk Web Scan'')}.\\
\hline

\textbf{Reporting (20)} &
Developers face reporting challenges when \gls{sst} tools do not present, export, deliver, or communicate vulnerability findings in a format needed for review or follow-up action. These challenges appear when developers cannot obtain accurate or complete report content, such as ZAP not showing the correct GET API in the generated report for security review documentation \textbf{(Q143~\cite{Q143}: ``zap not showing correct GET api in summery report'')}.\\
\hline

\textbf{False negatives (10)} &
Developers face false-negative challenges when \gls{sst} tools fail to detect vulnerabilities that developers expect the tool to identify during scanning, testing, or security analysis. These challenges appear when a scanner misses suspected code-level vulnerabilities, such as path traversal weaknesses in Scala code \textbf{(Q159~\cite{Q159}: ``Path traversal vulnerabilities not found at Scala code'')}.\\
\hline

\end{tabularx}
\end{table}

\begin{findingbox}
\textbf{Finding 2}. \textit{Developers face security finding interpretation and reliability challenges when \gls{sst} tools produce unclear explanations, false positives, incomplete reports, inconsistent results, or missed vulnerabilities that make security findings difficult to understand, verify, and trust.}
\end{findingbox}

\subsection{C2: Security Testing Guidance and Tool Selection}

C2 captures challenges developers face when they need to choose, compare, understand, or apply \gls{sst} tools and practices. Developers often work with different technologies, deployment environments, goals, and operational constraints. As a result, they need guidance on which tools to use, how tools differ, what features are available, and where reliable documentation or learning resources can be found. As shown in Table~\ref{tab:guidance-tool-selection-subtheme}, C2 includes five subcategories: tool suggestions, comparative evaluation, guidance, documentation/resources, and features.

\begin{table}[t]
\centering
\caption{C2 - Security Testing Guidance and Tool Selection}
\label{tab:guidance-tool-selection-subtheme}
\small
\setlength{\tabcolsep}{4pt}
\renewcommand{\arraystretch}{1.15}
\begin{tabularx}{\textwidth}{
>{\raggedright\arraybackslash}p{0.21\textwidth}
Y
}
\hline
\textbf{Name} & \textbf{Explanation} \\
\hline

\textbf{Guidance (56)} &
Developers face guidance challenges when they need practical direction for planning, scoping, or performing \gls{sst} activities across specific technologies, application features, protocols, or deployment environments. These challenges appear when developers ask how to test security-sensitive features, such as malicious file uploads \textbf{(Q73~\cite{Q73}: ``Security Testing - How to test file upload feature for malicious upload'')}.\\
\hline

\textbf{Tool suggestions (49)} &
Developers face tool-suggestion challenges when they need to identify suitable \gls{sst} tools for a specific platform, technology, protocol, testing goal, deployment environment, or project constraint. These challenges appear when developers look for penetration-testing tools for cloud environments, such as Google Cloud Platform \textbf{(Q40~\cite{Q40}: ``Are there any penetration test tools available for Google Cloud Platform?'')}.\\
\hline

\textbf{Documentation / Resources (41)} &
Developers face documentation and resource challenges when they need reliable references, examples, checklists, or learning materials to understand and apply \gls{sst} tools and practices. These challenges appear when developers look for structured resources to guide penetration-testing activities, such as network security testing checklists \textbf{(Q163~\cite{Q163}: ``How can I get a checklist to penetration testing of PS networks?'')}.\\
\hline

\textbf{Features (26)} &
Developers face feature-related challenges when they need to determine whether an \gls{sst} tool supports a capability, UI option, vulnerability display, or scan-control function. These challenges appear when developers cannot access tool features, such as a ZAP scan option unavailable for multi-selected URLs \textbf{(Q227~\cite{Q227}: ``OWASP Zap scan option is grayed-out for multi-selected URLs'')}.\\
\hline

\textbf{Comparative evaluation (9)} &
Developers face comparative-evaluation challenges when they need to compare \gls{sst} tools based on capabilities, coverage, cost, testing approach, integration support, workflow fit, or organizational requirements. These challenges appear when developers compare security testing tools for practical use, such as SOAP UI and Burp Suite \textbf{(Q13~\cite{Q13}: ``Security testing in soap ui or Burp suite?'')}.\\
\hline

\end{tabularx}
\end{table}

\begin{findingbox}
\textbf{Finding 3}. \textit{Developers face guidance and tool-selection challenges when they need clearer direction for selecting suitable \gls{sst} tools, comparing tool capabilities, understanding available features, or finding reliable resources for specific security testing tasks and application contexts.}
\end{findingbox}

\subsection{C3: Authentication and Authorization Testing}

C3 captures challenges developers face when protected application behavior must be configured, simulated, or verified during \gls{sst}. Security testing often requires authenticated users, roles, sessions, tokens, security contexts, filters, or external identity providers in the test environment. When these elements are missing or incorrectly configured, tests may fail, bypass protected behavior, or fail to represent realistic access-control conditions. As shown in Table~\ref{tab:authentication-authorization-subtheme}, C3 includes three subcategories: mocking, authentication functionality testing, and authentication test setup.

\begin{table}[t]
\centering
\caption{C3 - Authentication and Authorization Testing}
\label{tab:authentication-authorization-subtheme}
\small
\setlength{\tabcolsep}{4pt}
\renewcommand{\arraystretch}{1.15}
\begin{tabularx}{\textwidth}{
>{\raggedright\arraybackslash}p{0.21\textwidth}
Y
}
\hline
\textbf{Name} & \textbf{Explanation} \\
\hline

\textbf{Mocking (55)} &
Developers face mocking challenges when they need to simulate authenticated users, security contexts, sessions, roles, tokens, security managers, or external security components during unit and integration testing. These challenges appear when developers try to mock application-specific security identities or contexts, such as custom Spring Security users \textbf{(Q395~\cite{Q395}: ``Mock Custom User in Spring Security Test'')}.\\
\hline

\textbf{Authentication functionality testing (30)} &
Developers face authentication functionality testing challenges when they need to verify whether protected endpoints, authenticated workflows, role-based access, or security-enabled behavior works under test conditions in realistic application security contexts. These challenges appear when developers test applications that depend on identity providers or authentication servers, such as Keycloak-secured FastAPI endpoints \textbf{(Q540~\cite{Q540}: ``FastAPI Integration Testing with Keycloak User Authentication'')}.\\
\hline

\textbf{Authentication test setup (14)} &
Developers face authentication test setup challenges when they need to configure credentials, authentication flows, security filters, or test clients so that protected functionality can be tested correctly. These challenges appear when developers set up integration tests for applications protected by basic authentication, such as Spring Boot security tests requiring valid authenticated requests \textbf{(Q59~\cite{Q59}: ``How do I integration test Spring Boot with basic auth?'')}.\\
\hline

\end{tabularx}
\end{table}

\begin{findingbox}
\textbf{Finding 4}. \textit{Developers face authentication and authorization testing challenges when protected endpoints, authenticated users, security contexts, roles, tokens, or external identity providers must be configured, simulated, or verified during security testing in realistic application scenarios.}
\end{findingbox}

\subsection{C4: Security Tool Integration and Automation}

C4 captures challenges developers face when \gls{sst} tools are connected, migrated, or automated within development workflows involving CI/CD pipelines, testing frameworks, browsers, cloud platforms, IDEs, reporting systems, or authentication-dependent tasks. As shown in Table~\ref{tab:integration-automation-subtheme}, C4 includes three subcategories: integration, automation, and migration.

\begin{table}[t]
\centering
\caption{C4 - Security Tool Integration and Automation}
\label{tab:integration-automation-subtheme}
\small
\setlength{\tabcolsep}{4pt}
\renewcommand{\arraystretch}{1.15}
\begin{tabularx}{\textwidth}{
>{\raggedright\arraybackslash}p{0.21\textwidth}
Y
}
\hline
\textbf{Name} & \textbf{Explanation} \\
\hline

\textbf{Integration (65)} &
Developers face integration challenges when connecting \gls{sst} tools with development environments, CI/CD pipelines, browsers, testing frameworks, cloud platforms, IDEs, or reporting systems. These challenges appear when developers add security scans to CI/CD pipelines, such as Cloud Build vulnerability scanning \textbf{(Q435~\cite{Q435}: ``How to inlcude Container Registry vulnerability scans in the CI/CD script using Cloud Build on GCP'')}.\\
\hline

\textbf{Automation (12)} &
Developers face automation challenges when \gls{sst} activities are executed repeatedly, programmatically, or as part of a continuous testing workflow. These challenges appear when developers automate scanner operations, such as running ZAP fuzzing through automation interfaces \textbf{(Q424~\cite{Q424}: ``How to automate Fuzz operation in ZAP?'')}.\\
\hline

\textbf{Migration (9)} &
Developers face migration challenges when \gls{sst} rules, reports, tool configurations, executable artifacts, or workflows move from one tool, platform, version, or runtime environment to another. These challenges appear when developers migrate static-analysis rules across tools, such as moving FindBugs rules into SonarQube/Squid \textbf{(Q76~\cite{Q76}: ``How to migrate FindBugs rule EI\_EXPOSE\_REP to Squid?'')}.\\
\hline

\end{tabularx}
\end{table}

\begin{findingbox}
\textbf{Finding 5}. \textit{Developers face integration and automation challenges when \gls{sst} tools are connected, migrated, or automated within existing development environments, CI/CD pipelines, testing frameworks, cloud platforms, operational workflows, and evolving deployment contexts.}
\end{findingbox}

\subsection{C5: Security Test Configuration and Environment Setup}

C5 captures challenges developers face when \gls{sst} tools and test environments are configured before security testing can run correctly. These challenges arise because security testing often depends on tool settings, security framework configuration, scan contexts, proxies, authentication mechanisms, fuzzing inputs, and runtime conditions. When these elements are missing or misconfigured, security tests may fail, skip protected behavior, or produce unreliable results. As shown in Table~\ref{tab:security-test-configuration-environment-subtheme}, C5 includes three subcategories: configuration, test environment readiness, and fuzzing setup.

\begin{table}[t]
\centering
\caption{C5 - Security Test Configuration and Environment Setup}
\label{tab:security-test-configuration-environment-subtheme}
\small
\setlength{\tabcolsep}{4pt}
\renewcommand{\arraystretch}{1.15}
\begin{tabularx}{\textwidth}{
>{\raggedright\arraybackslash}p{0.21\textwidth}
Y
}
\hline
\textbf{Name} & \textbf{Explanation} \\
\hline

\textbf{Configuration (53)} &
Developers face configuration challenges when they need to set up \gls{sst} tools, security frameworks, quality gates, proxies, scan contexts, or test settings so that security testing behaves as expected. These challenges appear when developers struggle with initial tool or framework setup, such as configuring ZAP for Vaadin applications \textbf{(Q77~\cite{Q77}: ``Zap Vaadin setup issues'')}.\\
\hline

\textbf{Test environment readiness (7)} &
Developers face test environment readiness challenges when the application context, security configuration, runtime state, or test setup is not prepared for executing \gls{sst} activities. These challenges appear when developers need to load security for selected tests while keeping other tests independent from the security setup \textbf{(Q416~\cite{Q416}: ``Loading security for some specific tests only'')}.\\
\hline

\textbf{Fuzzing setup (5)} &
Developers face fuzzing setup challenges when they need to prepare fuzzing tools, inputs, tokens, or test harnesses so that fuzzing reaches the security target. These challenges appear when developers configure fuzzers for protected APIs, such as adding bearer-token authentication to RESTler API fuzzing \textbf{(Q320~\cite{Q320}: ``RESTler API Fuzzing Add Authentication Bearer token'')}.\\
\hline

\end{tabularx}
\end{table}

\begin{findingbox}
\textbf{Finding 6}. \textit{Developers face security test configuration and environment setup challenges when \gls{sst} tools, security frameworks, test contexts, proxies, authentication mechanisms, or fuzzing setup conditions are not configured for reliable security testing execution.}
\end{findingbox}

\subsection{C6: Security Rule, Policy, and Coverage Validation}

C6 captures challenges developers face when security rules, policies, test cases, or scanner outputs are mapped, customized, reproduced, validated, or assessed for coverage. Security testing results become useful when developers can identify which rule produced a finding, whether expected security behavior is covered, and whether a rule or policy is enforced in the target context. As shown in Table~\ref{tab:rule-policy-coverage-subtheme}, C6 includes five subcategories: validation, coverage, customization, reproducibility, and mapping across rule-oriented security testing workflows.

\begin{table}[t]
\centering
\caption{C6 - Security Rule, Policy, and Coverage Validation}
\label{tab:rule-policy-coverage-subtheme}
\small
\setlength{\tabcolsep}{4pt}
\renewcommand{\arraystretch}{1.15}
\begin{tabularx}{\textwidth}{
>{\raggedright\arraybackslash}p{0.21\textwidth}
Y
}
\hline
\textbf{Name} & \textbf{Explanation} \\
\hline

\textbf{Validation (22)} &
Developers face validation challenges when they need to confirm whether security rules, certificates, policies, or validation mechanisms are enforced during testing or runtime checks. These challenges appear when developers encounter errors while validating security-related artifacts, such as certificate paths \textbf{(Q346~\cite{Q346}: ``Error when validating certificate path'')}.\\
\hline

\textbf{Coverage (13)} &
Developers face coverage challenges when they need to determine whether \gls{sst} tools, rules, or reports cover specific vulnerability categories, standards, URLs, APIs, or files. These challenges appear when developers try to understand how scanner alerts map to broader security standards, such as OWASP \textbf{(Q230~\cite{Q230}: ``ZAP alert categorization in owasp top 10 vulnerabilities'')}.\\
\hline

\textbf{Customization (10)} &
Developers face customization challenges when they need to create, modify, enable, or adapt \gls{sst} rules and tool behavior for project-specific security analysis. These challenges appear when developers try to make custom security rules work for specific platforms, such as applying Java custom rules to an Android project in SonarQube \textbf{(Q87~\cite{Q87}: ``Can I make Java Custom rules analyze a Android Project on SonarQube?'')}.\\
\hline

\textbf{Reproducibility (6)} &
Developers face reproducibility challenges when they need to recreate, confirm, or demonstrate a reported issue under controlled testing conditions. These challenges appear when developers try to inject controlled events or failures to test recovery behavior at a specific point in code \textbf{(Q32~\cite{Q32}: ``How to simulate an external event during testing at a particular point in code'')}.\\
\hline

\textbf{Mapping (4)} &
Developers face mapping challenges when they need to connect security rules, findings, or tool outputs to rule definitions, categories, code locations, analysis configurations, or activation settings. These challenges appear when developers try to identify which activated SonarQube rule produced an \texttt{ArrayOutOfBoundsException} after enabling multiple rules \textbf{(Q365~\cite{Q365}: ``How to find the rule throwing an ArrayOutOfBoundsException after activating a bunch of rules in Sonarqube?'')}.\\
\hline

\end{tabularx}
\end{table}

\begin{findingbox}
\textbf{Finding 7}. \textit{Developers face rule, policy, and coverage validation challenges when \gls{sst} tools do not clearly show which rules apply, what behavior is covered, whether custom rules work correctly, or whether reported issues can be reproduced and validated in the target context.}
\end{findingbox}

\subsection{C7: Remediation, Dependency, and Supply Chain Actionability}

C7 captures challenges developers face when \gls{sst} findings involve dependencies, container images, packages, source-code locations, or remediation steps. Security findings become actionable when developers identify the affected artifact, understand the dependency path or code location, and determine a feasible fix. As shown in Table~\ref{tab:remediation-dependency-supplychain-subtheme}, C7 includes three subcategories: remediation actionability, dependency and supply chain vulnerability, and vulnerability localization.

\begin{table}[t]
\centering
\caption{C7 - Remediation, Dependency, and Supply Chain Actionability}
\label{tab:remediation-dependency-supplychain-subtheme}
\small
\setlength{\tabcolsep}{4pt}
\renewcommand{\arraystretch}{1.15}
\begin{tabularx}{\textwidth}{
>{\raggedright\arraybackslash}p{0.21\textwidth}
Y
}
\hline
\textbf{Name} & \textbf{Explanation} \\
\hline

\textbf{Remediation Actionability (19)} &
Developers face remediation actionability challenges when \gls{sst} findings do not provide a clear, feasible, or directly applicable path for fixing the reported vulnerability. These challenges appear when scanner-recommended fixes cannot be applied because no compatible, patched, or fixed dependency version is available \textbf{(Q297~\cite{Q297}: ``Jfrog Xray Violations should not occure for libraries where there is no fix version'')}.\\
\hline

\textbf{Dependency and Supply Chain Vulnerability (16)} &
Developers face dependency and supply chain vulnerability challenges when \gls{sst} activities involve identifying, understanding, or testing vulnerabilities in third-party libraries, package-lock files, container images, or deployed software artifacts. These challenges appear when developers need to scan deployed containers or runtime artifacts for known vulnerabilities after release \textbf{(Q127~\cite{Q127}: ``How to automatically scan deployed containers for security vulnerabilities?'')}.\\
\hline

\textbf{Vulnerability Localization (9)} &
Developers face vulnerability localization challenges when they need to trace a reported security issue, vulnerable function, package, or dependency artifact back to the source-code location or dependency path that causes the finding. These challenges appear when developers connect tool outputs to source-code elements, such as identifying function names reported by Xcode's \texttt{otool} output \textbf{(Q270~\cite{Q270}: ``How to find out in source code the function names the Xcode's otool outputs for you?'')}.\\
\hline

\end{tabularx}
\end{table}

\begin{findingbox}
\textbf{Finding 8}. \textit{Developers face remediation, dependency, and supply chain actionability challenges when \gls{sst} tools report vulnerable packages, container images, dependency artifacts, or code-level issues without clearly identifying the affected source, dependency path, practical impact, or feasible remediation.}
\end{findingbox}

\subsection{C8: Security Risks, Constraints, and Performance in Testing Practice}

C8 captures challenges developers face when \gls{sst} activities introduce additional risks or operational burdens during testing. Security testing may expose sensitive information, require careful handling of credentials or tokens, increase execution time, or raise legal and authorization concerns in real-world developer testing environments. As shown in Table~\ref{tab:security-risks-constraints-performance-subtheme}, C8 includes four subcategories: performance, secret management, information leakage, and legal constraints.

\begin{table}[t]
\centering
\caption{C8 - Security Risks, Constraints, and Performance in Testing Practice}
\label{tab:security-risks-constraints-performance-subtheme}
\small
\setlength{\tabcolsep}{4pt}
\renewcommand{\arraystretch}{1.15}
\begin{tabularx}{\textwidth}{
>{\raggedright\arraybackslash}p{0.21\textwidth}
Y
}
\hline
\textbf{Name} & \textbf{Explanation} \\
\hline

\textbf{Performance (11)} &
Developers face performance challenges when \gls{sst} activities introduce long execution time, high resource consumption, or noisy tool behavior that slows testing workflows. These challenges appear when developers report that security scanners take too long to complete, such as ZAP scans becoming time consuming during full application security assessment \textbf{(Q22~\cite{Q22}: ``ZAP security tool is very time consuming'')}.\\
\hline

\textbf{Secret management (8)} &
Developers face secret management challenges when \gls{sst} activities involve handling, protecting, correlating, or detecting tokens, passwords, and authentication secrets. These challenges appear when developers need to manage session tokens during security testing, such as correlating Temenos T24 session tokens in JMeter across repeated requests \textbf{(Q326~\cite{Q326}: ``How to Handle Security Violations and Correlate Session Tokens in JMeter for Temenos T24?'')}.\\
\hline

\textbf{Information leakage (5)} &
Developers face information leakage challenges when \gls{sst} activities may expose sensitive data, credentials, secrets, configuration details, vulnerability information, or security-related metadata. These challenges appear when developers worry that security testing tools may capture or reveal sensitive application data during proxy-based testing and traffic inspection \textbf{(Q129~\cite{Q129}: ``Sensitive data exposure in Burp'')}.\\
\hline

\textbf{Legal constraints (4)} &
Developers face legal constraint challenges when \gls{sst} activities raise concerns about authorization, data ownership, third-party service rules, or the legality of automated security testing behavior. These challenges appear when developers question whether source code can be safely uploaded to cloud-based security scanning services \textbf{(Q203~\cite{Q203}: ``Is it safe to upload code to Fortify on demand for scanning?'')}.\\
\hline

\end{tabularx}
\end{table}

\begin{findingbox}
\textbf{Finding 9}. \textit{Developers face security risks, constraints, and performance challenges when \gls{sst} activities expose sensitive information, require secure handling of credentials or tokens, slow testing workflows, or raise legal and authorization concerns during security testing practice.}
\end{findingbox}

Overall, as shown in Fig.~\ref{fig:sst-taxonomy}, the taxonomy indicates that \gls{sst} challenges span the security testing workflow, from tool selection and configuration to finding interpretation, dependency analysis, and remediation. This distribution suggests that \gls{sst} is not only vulnerability detection but also a workflow, interpretation, and actionability challenge for developers.

\section{Findings from RQ2: Problem Characteristics of SST Challenges}
\label{sec:rq2_popularity_difficulty}

This section presents the quantitative findings for RQ2 by analyzing \gls{sst} challenges in terms of prevalence, difficulty, correlation, temporal trends, and co-occurrence.

\subsection{Answer to RQ2.1: Which \gls{sst} challenges are most prevalent in the developer community?}

Table~\ref{tab:sst_popularity_analysis} reports the prevalence metrics for \gls{sst} categories and subcategories. At the category level, \textit{Authentication and Authorization Testing} receives the highest average views (3,803.59) and average score (2.66), indicating strong community attention to authentication-related testing problems. It is followed in average views by \textit{Security Finding Interpretation and Reliability} (2,407.38), \textit{Security Test Configuration and Environment Setup} (2,392.83), and \textit{Security Testing Guidance and Tool Selection} (2,373.30). \textit{Security Testing Guidance and Tool Selection} has the highest average answers (1.24), suggesting stronger community response to tool-selection and guidance questions.

At the subcategory level, \textit{Mocking} receives the highest average views (4,391.35), followed by \textit{Explainability} (3,900.66), \textit{Authentication test setup} (3,603.14), \textit{Tool suggestions} (3,492.86), \textit{Vulnerability localization} (3,328.33), and \textit{Validation} (3,327.41). Comment activity is highest for challenges involving contextual clarification, including \textit{Information leakage} (2.80), \textit{Legal constraints} (2.75), \textit{False positives} (2.18), \textit{Explainability} (2.15), and \textit{Vulnerability localization} (2.11). These results show that developers frequently seek help with authentication simulation, scanner-output interpretation, tool selection, validation, and vulnerability tracing, while sensitive, legal, and interpretation-heavy questions often require more discussion.

\begin{findingbox}
\textbf{Finding 10}. \textit{Authentication and authorization testing is the most visible \gls{sst} challenge category, with the highest average views (3,803.59) and average score (2.66). At the subcategory level, mocking receives the highest visibility (4,391.35 average views), while explainability, tool suggestions, validation, and vulnerability localization also attract substantial developer attention.}
\end{findingbox}

\begin{table}[!t]
\centering
\caption{Prevalence Metrics for \gls{sst} Categories and Subcategories}
\label{tab:sst_popularity_analysis}
\scriptsize
\setlength{\tabcolsep}{3pt}
\renewcommand{\arraystretch}{1.08}
\begin{tabularx}{\columnwidth}{
>{\raggedright\arraybackslash}X
>{\raggedleft\arraybackslash}p{0.13\columnwidth}
>{\raggedleft\arraybackslash}p{0.13\columnwidth}
>{\raggedleft\arraybackslash}p{0.13\columnwidth}
>{\raggedleft\arraybackslash}p{0.13\columnwidth}
}
\toprule
\textbf{Category / Subcategory} & \textbf{Avg. View} & \textbf{Avg. Comment} & \textbf{Avg. Answer} & \textbf{Avg. Score} \\
\midrule

\textbf{C1: Security Finding Interpretation and Reliability} & \textbf{2407.38} & \textbf{1.79} & \textbf{0.94} & \textbf{1.69} \\
\quad False positives & 1953.45 & 2.18 & 0.79 & 1.64 \\
\quad Explainability & 3900.66 & 2.15 & 0.98 & 2.35 \\
\quad Inconsistency & 1882.77 & 1.71 & 1.00 & 0.94 \\
\quad Reporting & 1669.00 & 1.80 & 1.05 & 2.55 \\
\quad False negatives & 1294.60 & 0.50 & 0.90 & 0.90 \\
\addlinespace[2pt]

\textbf{C2: Security Testing Guidance and Tool Selection} & \textbf{2373.30} & \textbf{1.08} & \textbf{1.24} & \textbf{1.04} \\
\quad Guidance & 2064.43 & 1.54 & 1.23 & 1.12 \\
\quad Tool suggestions & 3492.86 & 0.65 & 1.39 & 1.69 \\
\quad Documentation / Resources & 2534.66 & 0.80 & 1.12 & 0.71 \\
\quad Features & 1742.38 & 1.23 & 1.08 & 0.62 \\
\quad Comparative evaluation & 2868.11 & 0.44 & 1.33 & 2.78 \\
\addlinespace[2pt]

\textbf{C3: Authentication and Authorization Testing} & \textbf{3803.59} & \textbf{1.50} & \textbf{1.14} & \textbf{2.66} \\
\quad Mocking & 4391.35 & 1.78 & 1.15 & 3.00 \\
\quad Authentication functionality testing & 2483.80 & 1.43 & 1.07 & 1.57 \\
\quad Authentication test setup & 3603.14 & 0.71 & 1.14 & 3.14 \\
\addlinespace[2pt]

\textbf{C4: Security Tool Integration and Automation} & \textbf{1817.51} & \textbf{1.29} & \textbf{1.05} & \textbf{1.36} \\
\quad Integration & 2162.85 & 1.31 & 1.11 & 1.43 \\
\quad Automation & 520.83 & 0.83 & 0.75 & 0.67 \\
\quad Migration & 870.56 & 1.78 & 1.00 & 1.67 \\
\addlinespace[2pt]

\textbf{C5: Security Test Configuration and Environment Setup} & \textbf{2392.83} & \textbf{1.45} & \textbf{1.13} & \textbf{2.20} \\
\quad Configuration & 2377.28 & 1.40 & 1.19 & 2.34 \\
\quad Test environment readiness & 2510.57 & 1.86 & 0.71 & 1.14 \\
\quad Fuzzing setup & 298.60 & 0.00 & 0.80 & 1.00 \\
\addlinespace[2pt]

\textbf{C6: Security Rule, Policy, and Coverage Validation} & \textbf{1936.91} & \textbf{1.21} & \textbf{1.04} & \textbf{2.02} \\
\quad Validation & 3327.41 & 1.41 & 1.18 & 3.50 \\
\quad Coverage & 1026.23 & 0.38 & 1.23 & 0.77 \\
\quad Customization & 1061.20 & 1.70 & 0.90 & 1.50 \\
\quad Reproducibility & 473.83 & 1.00 & 0.83 & 0.33 \\
\quad Mapping & 893.75 & 1.25 & 0.50 & 0.75 \\
\addlinespace[2pt]

\textbf{C7: Remediation, Dependency, and Supply Chain Actionability} & \textbf{1992.43} & \textbf{1.57} & \textbf{1.00} & \textbf{1.74} \\
\quad Remediation actionability & 1942.47 & 1.63 & 1.11 & 1.16 \\
\quad Dependency and supply chain vulnerability & 1218.12 & 1.12 & 0.94 & 1.81 \\
\quad Vulnerability localization & 3328.33 & 2.11 & 0.89 & 2.78 \\
\addlinespace[2pt]

\textbf{C8: Security Risks, Constraints, and Performance in Testing Practice} & \textbf{2097.39} & \textbf{1.82} & \textbf{1.04} & \textbf{0.75} \\
\quad Performance & 3238.64 & 1.45 & 1.45 & 1.55 \\
\quad Secret management & 1591.75 & 1.25 & 0.75 & 0.12 \\
\quad Information leakage & 1610.00 & 2.80 & 0.80 & 0.80 \\
\quad Legal constraints & 579.50 & 2.75 & 0.75 & -0.25 \\

\bottomrule
\end{tabularx}
\end{table}

\subsection{Answer to RQ2.2: Which \gls{sst} challenge categories exhibit more difficulty in terms of obtaining satisfactory resolutions?}

Table~\ref{tab:security_metrics_full} reports the difficulty metrics for \gls{sst} categories and subcategories using answer availability, accepted-answer rate, response time, and question size. At the category level, \textit{Security Rule, Policy, and Coverage Validation} appears among the most difficult categories, with a low \textbf{AnswerRate} (0.77), low \textbf{AcceptRate} (0.34), the longest \textbf{TimeAcceptedAnswer} (41.66 hours), and the largest \textbf{TextSize} (2,631.75 characters). \textit{Security Test Configuration and Environment Setup} shows difficulty, with a low \textbf{AnswerRate} (0.77), low \textbf{AcceptRate} (0.33), and the longest \textbf{TimeFirstAnswer} (48.45 hours). \textit{Security Finding Interpretation and Reliability} shows similar difficulty, with a low \textbf{AnswerRate} (0.78), low \textbf{AcceptRate} (0.32), and long \textbf{TimeFirstAnswer} (25.28 hours).

At the subcategory level, \textit{Inconsistency} has a low \textbf{AcceptRate} (0.19) and long \textbf{TimeFirstAnswer} (154.87 hours), indicating difficulty in resolving inconsistent tool behavior. \textit{Test environment readiness} also appears difficult, with a low \textbf{AnswerRate} (0.57) and \textbf{AcceptRate} (0.14), while \textit{Vulnerability localization} has a low \textbf{AnswerRate} (0.67) and the lowest \textbf{AcceptRate} in its category (0.11). \textit{False positives} also shows difficulty, with a low \textbf{AnswerRate} (0.71) and \textbf{AcceptRate} (0.29). In contrast, \textit{Authentication and Authorization Testing} appears comparatively easier to resolve, with the highest category-level \textbf{AcceptRate} (0.52), a high \textbf{AnswerRate} (0.84), and short response times. Its \textit{Authentication test setup} subcategory has the highest \textbf{AcceptRate} (0.64).

\begin{findingbox}
\textbf{Finding 11}. \textit{Security rule, policy, and coverage validation is one of the most difficult \gls{sst} challenge categories, with the longest accepted-answer time (41.66 hours) and the largest average question size (2,631.75 characters). Security test configuration and environment setup also shows high difficulty, with the longest first-answer time (48.45 hours), while authentication and authorization testing appears comparatively easier to resolve due to the highest accepted-answer rate (0.52).}
\end{findingbox}

\begin{table}[!t]
\centering
\caption{Difficulty Metrics for \gls{sst} Categories and Subcategories}
\label{tab:security_metrics_full}
\scriptsize
\setlength{\tabcolsep}{2pt}
\renewcommand{\arraystretch}{1.08}
\begin{tabularx}{\columnwidth}{
>{\raggedright\arraybackslash}X
>{\raggedleft\arraybackslash}p{0.10\columnwidth}
>{\raggedleft\arraybackslash}p{0.10\columnwidth}
>{\raggedleft\arraybackslash}p{0.12\columnwidth}
>{\raggedleft\arraybackslash}p{0.12\columnwidth}
>{\raggedleft\arraybackslash}p{0.12\columnwidth}
}
\toprule
\textbf{Category / Subcategory} & \textbf{Ans. Rate} & \textbf{Acc. Rate} & \textbf{First Ans. (h)} & \textbf{Acc. Ans. (h)} & \textbf{Text Size} \\
\midrule

\textbf{C1: Security Finding Interpretation and Reliability} & \textbf{0.78} & \textbf{0.32} & \textbf{25.28} & \textbf{14.44} & \textbf{1516.85} \\
\quad False positives & 0.71 & 0.29 & 16.20 & 4.71 & 1424.34 \\
\quad Explainability & 0.78 & 0.31 & 23.24 & 16.87 & 1542.58 \\
\quad Inconsistency & 0.77 & 0.19 & 154.87 & 132.90 & 1194.16 \\
\quad Reporting & 0.90 & 0.50 & 24.74 & 46.14 & 2406.20 \\
\quad False negatives & 0.70 & 0.40 & 62.94 & 55.88 & 1922.90 \\
\addlinespace[2pt]

\textbf{C2: Security Testing Guidance and Tool Selection} & \textbf{0.89} & \textbf{0.38} & \textbf{8.04} & \textbf{4.58} & \textbf{845.48} \\
\quad Guidance & 0.86 & 0.36 & 3.02 & 4.46 & 821.64 \\
\quad Tool suggestions & 0.92 & 0.31 & 12.21 & 6.12 & 950.49 \\
\quad Documentation / Resources & 0.88 & 0.41 & 20.28 & 2.14 & 1535.12 \\
\quad Features & 0.92 & 0.50 & 5.44 & 13.97 & 471.73 \\
\quad Comparative evaluation & 1.00 & 0.22 & 16.77 & 15.91 & 600.89 \\
\addlinespace[2pt]

\textbf{C3: Authentication and Authorization Testing} & \textbf{0.84} & \textbf{0.52} & \textbf{3.85} & \textbf{4.85} & \textbf{2572.57} \\
\quad Mocking & 0.82 & 0.51 & 2.90 & 4.18 & 2669.55 \\
\quad Authentication functionality testing & 0.87 & 0.47 & 5.54 & 4.50 & 2641.77 \\
\quad Authentication test setup & 0.86 & 0.64 & 2.37 & 20.06 & 1947.00 \\
\addlinespace[2pt]

\textbf{C4: Security Tool Integration and Automation} & \textbf{0.82} & \textbf{0.38} & \textbf{9.50} & \textbf{8.56} & \textbf{1588.73} \\
\quad Integration & 0.85 & 0.35 & 13.48 & 10.36 & 1744.52 \\
\quad Automation & 0.67 & 0.42 & 3.11 & 3.78 & 933.58 \\
\quad Migration & 0.89 & 0.44 & 10.76 & 5.51 & 1244.22 \\
\addlinespace[2pt]

\textbf{C5: Security Test Configuration and Environment Setup} & \textbf{0.77} & \textbf{0.33} & \textbf{48.45} & \textbf{14.04} & \textbf{2296.52} \\
\quad Configuration & 0.79 & 0.36 & 48.45 & 17.46 & 2461.85 \\
\quad Test environment readiness & 0.57 & 0.14 & 86.12 & 3.92 & 1044.71 \\
\quad Fuzzing setup & 0.80 & 0.40 & 0.28 & 0.27 & 1242.00 \\
\addlinespace[2pt]

\textbf{C6: Security Rule, Policy, and Coverage Validation} & \textbf{0.77} & \textbf{0.34} & \textbf{16.77} & \textbf{41.66} & \textbf{2631.75} \\
\quad Validation & 0.77 & 0.32 & 31.55 & 144.49 & 2670.36 \\
\quad Coverage & 1.00 & 0.54 & 14.35 & 20.93 & 2919.15 \\
\quad Customization & 0.70 & 0.20 & 11.26 & 8.65 & 1478.20 \\
\quad Reproducibility & 0.67 & 0.33 & 87.20 & 4795.48 & 1215.83 \\
\quad Mapping & 0.50 & 0.25 & 27010.45 & 4254.54 & 5456.50 \\
\addlinespace[2pt]

\textbf{C7: Remediation, Dependency, and Supply Chain Actionability} & \textbf{0.81} & \textbf{0.33} & \textbf{6.25} & \textbf{4.71} & \textbf{1825.45} \\
\quad Remediation actionability & 0.89 & 0.26 & 78.81 & 78.81 & 1964.68 \\
\quad Dependency and supply chain vulnerability & 0.81 & 0.50 & 0.90 & 0.74 & 2559.88 \\
\quad Vulnerability localization & 0.67 & 0.11 & 3.21 & 68.99 & 2544.78 \\
\addlinespace[2pt]

\textbf{C8: Security Risks, Constraints, and Performance in Testing Practice} & \textbf{0.79} & \textbf{0.32} & \textbf{4.83} & \textbf{48.85} & \textbf{1086.04} \\
\quad Performance & 1.00 & 0.36 & 3.64 & 26.04 & 1502.82 \\
\quad Secret management & 0.62 & 0.50 & 1.25 & 224.60 & 963.12 \\
\quad Information leakage & 0.60 & 0.20 & 693.00 & 693.00 & 760.00 \\
\quad Legal constraints & 0.75 & 0.00 & 0.27 & -- & 593.25 \\

\bottomrule
\end{tabularx}
\begin{flushleft}
\footnotesize
Note: ``--'' indicates that no accepted answer was available for computing the accepted-answer time.
\end{flushleft}
\end{table}

\subsection{Answer to RQ2.3: What is the relationship between the prevalence of a challenge and its perceived technical difficulty?}

\begin{figure*}[t]
    \centering
    \includegraphics[width=\textwidth]{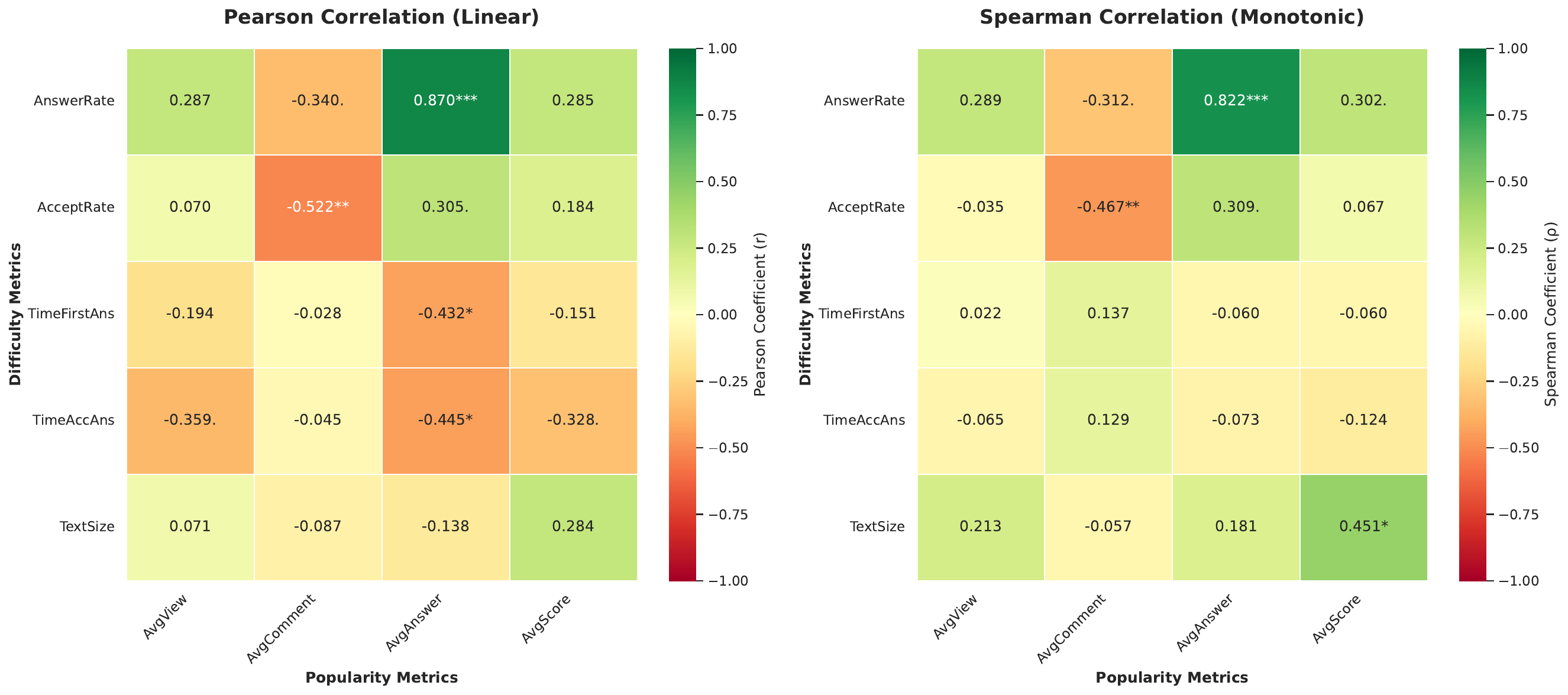}
    \caption{Comparison of Pearson and Spearman correlation matrices between difficulty and prevalence metrics. Significance levels: $p < 0.10 (.)$, $p < 0.05 (*)$, $p < 0.01 (**)$, $p < 0.001 (***)$.}
    \label{fig:correlation_heatmaps}
\end{figure*}

We examine how prevalence and difficulty metrics relate to each other using correlation analysis, with results shown in Figure~\ref{fig:correlation_heatmaps}.

The Pearson results in Fig.~\ref{fig:correlation_heatmaps} (left) show a strong positive relationship between \textbf{AnswerRate} and \textbf{avgAnswer} ($r = 0.870$, $p < 0.001$), indicating that categories with more answered questions also receive more answers on average. Response-time metrics are negatively correlated with answer volume: \textbf{TimeFirstAnswer} with \textbf{avgAnswer} ($r = -0.432$, $p < 0.05$) and \textbf{TimeAcceptedAnswer} with \textbf{avgAnswer} ($r = -0.445$, $p < 0.05$). These results suggest that faster-resolved categories tend to receive stronger answer activity. In addition, \textbf{AcceptRate} is negatively correlated with \textbf{avgComment} ($r = -0.522$, $p < 0.01$), indicating that questions with lower satisfactory resolution rates may require more comment-based clarification.

The Spearman results in Fig.~\ref{fig:correlation_heatmaps} (right) confirm the strongest Pearson relationship: \textbf{AnswerRate} remains strongly correlated with \textbf{avgAnswer} ($\rho = 0.822$, $p < 0.001$). \textbf{AcceptRate} is also negatively correlated with \textbf{avgComment} ($\rho = -0.467$, $p < 0.01$), suggesting that categories with lower accepted-answer rates tend to involve more clarification and discussion. The response-time correlations are weaker and not statistically significant in the Spearman analysis, indicating that response delays do not follow a stable rank-based relationship with prevalence metrics. Overall, \textbf{AnswerRate} and \textbf{avgAnswer} show the most stable relationship across both tests, \textbf{AcceptRate} and \textbf{avgComment} show an inverse relationship, and \textbf{TextSize} has a positive Spearman correlation with \textbf{avgScore} ($\rho = 0.451$, $p < 0.05$). These results show that prevalence and difficulty are related but not uniformly: answer-related metrics align with community response, while comment activity reflects clarification needs rather than direct popularity.

\begin{findingbox}
\textbf{Finding 12}. \textit{Answer availability is associated with community response in \gls{sst} discussions. AnswerRate has a strong positive correlation with avgAnswer in both Pearson ($r = 0.870$, $p < 0.001$) and Spearman ($\rho = 0.822$, $p < 0.001$) analyses. In contrast, AcceptRate is negatively correlated with avgComment, suggesting that less conclusively resolved challenges require more clarification and discussion.}
\end{findingbox}

\subsection{Answer to RQ2.4: How do \gls{sst} challenge categories trend and evolve over time?}

Table~\ref{tab:temporal_trends_category} reports the temporal trend results for the \gls{sst} challenge categories. Developer attention increases for \textit{Security Finding Interpretation and Reliability} ($p=0.002$) and \textit{Remediation, Dependency, and Supply Chain Actionability} ($p=0.006$), indicating growing attention to scanner-output interpretation, false positives, inconsistent findings, vulnerable dependencies, affected components, and actionable remediation guidance in evolving security testing workflows today.

In contrast, \textit{Security Testing Guidance and Tool Selection} ($p=0.004$) and \textit{Authentication and Authorization Testing} ($p=0.033$) show decreasing trends relative to other \gls{sst} categories. The remaining categories remain consistent over time, suggesting that integration, configuration, validation, performance, and testing-risk concerns continue to be persistent parts of developer-facing \gls{sst} practice across real-world tool-supported development environments.

\begin{table}[!t]
\centering
\caption{Temporal Trend Analysis of \gls{sst} Challenge Categories}
\label{tab:temporal_trends_category}
\small
\setlength{\tabcolsep}{5pt}
\renewcommand{\arraystretch}{1.15}
\begin{tabularx}{\textwidth}{
>{\raggedright\arraybackslash}p{0.48\textwidth}
>{\centering\arraybackslash}p{0.13\textwidth}
>{\centering\arraybackslash}p{0.18\textwidth}
>{\centering\arraybackslash}p{0.14\textwidth}
}
\toprule
\textbf{Category} & \textbf{Qs. Prop.} & \textbf{Cox-Stuart} & \textbf{Trend} \\
\midrule
C1: Security Finding Interpretation and Reliability & 24.93\% & $\Uparrow$, $p=0.002$ & \textbf{Increasing} \\
C2: Security Testing Guidance and Tool Selection & 23.51\% & $\Downarrow$, $p=0.004$ & \textbf{Decreasing} \\
C3: Authentication and Authorization Testing & 13.60\% & $\Downarrow$, $p=0.033$ & \textbf{Decreasing} \\
C4: Security Tool Integration and Automation & 12.04\% & $\Downarrow$, $p=0.878$ & Consistent \\
C5: Security Test Configuration and Environment Setup & 8.50\% & $\Downarrow$, $p=0.743$ & Consistent \\
C6: Security Rule, Policy, and Coverage Validation & 7.51\% & $\Downarrow$, $p=0.110$ & Consistent \\
C7: Remediation, Dependency, and Supply Chain Actionability & 5.95\% & $\Uparrow$, $p=0.006$ & \textbf{Increasing} \\
C8: Security Risks, Constraints, and Performance in Testing Practice & 3.97\% & $\Downarrow$, $p=0.678$ & Consistent \\
\bottomrule
\end{tabularx}
\end{table}

\begin{findingbox}
\textbf{Finding 13}. \textit{Developer attention is shifting toward security finding interpretation and remediation-related actionability. Security finding interpretation and reliability shows a significant increasing trend ($p=0.002$), and remediation, dependency, and supply chain actionability also increases significantly ($p=0.006$). In contrast, guidance/tool selection and authentication/authorization testing show decreasing trends, while the remaining categories remain consistent over time.}
\end{findingbox}

\subsection{Answer to RQ2.5: Which \gls{sst} challenge subcategories co-occur in developer questions?}

\begin{figure*}[t]
    \centering
    \begin{subfigure}[b]{0.48\textwidth}
        \centering
        \includegraphics[width=\textwidth]{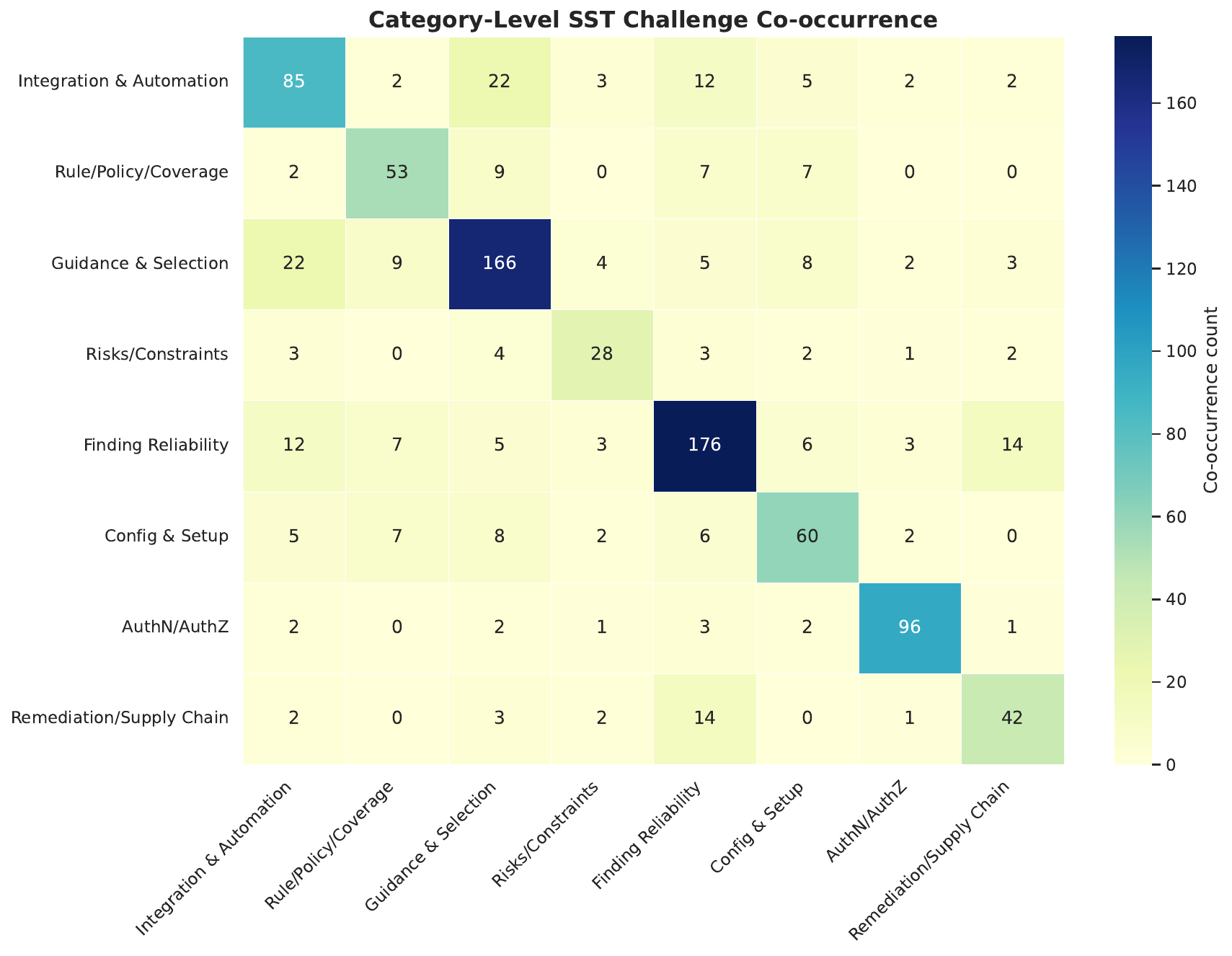}
        \caption{Category-level co-occurrence}
        \label{fig:category_cooccurrence_heatmap}
    \end{subfigure}
    \hfill
    \begin{subfigure}[b]{0.48\textwidth}
        \centering
        \includegraphics[width=\textwidth]{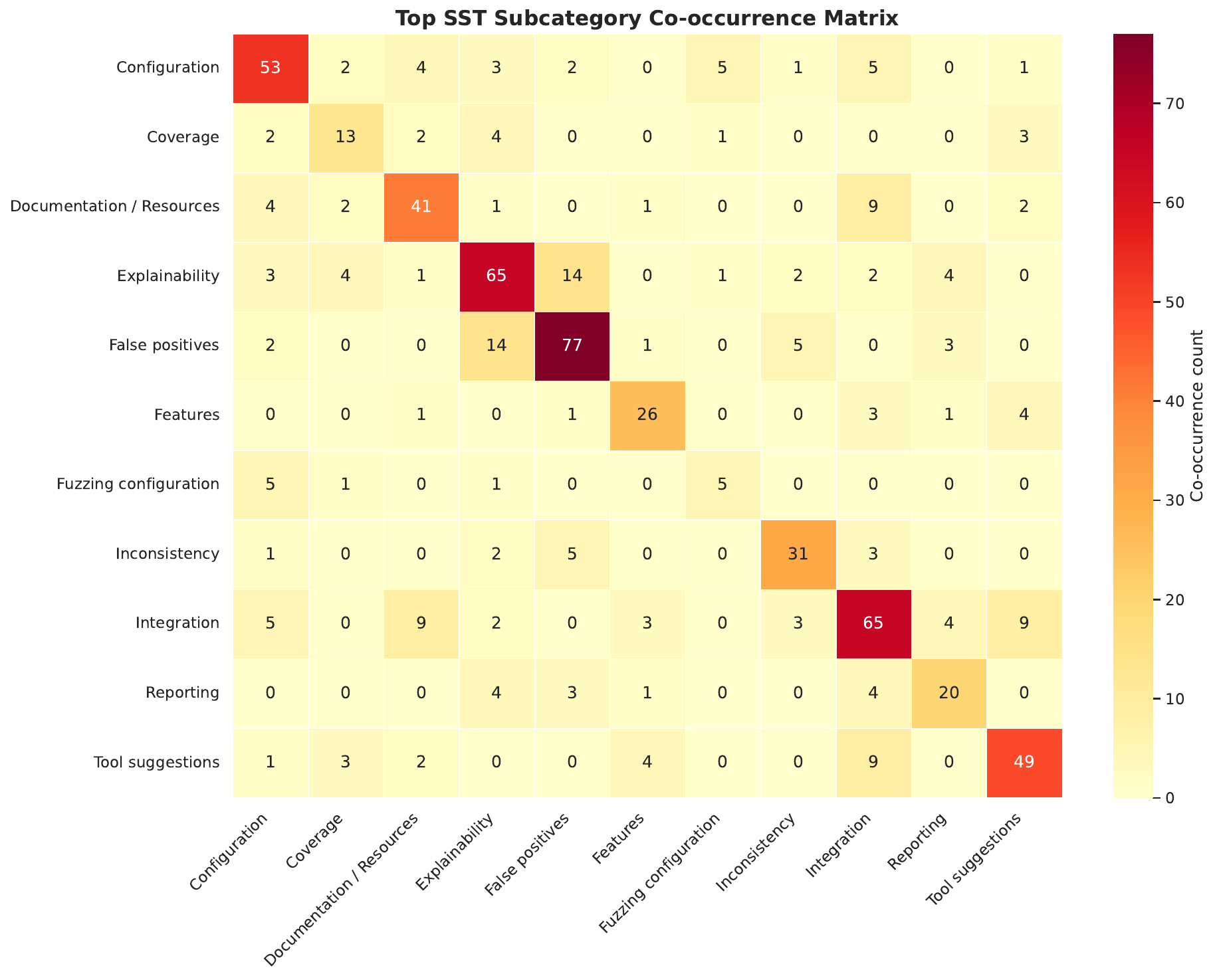}
        \caption{Subcategory-level co-occurrence}
        \label{fig:subcategory_cooccurrence_heatmap}
    \end{subfigure}
    
    \caption{Co-occurrence patterns among \gls{sst} challenges at category and subcategory levels. Darker cells indicate stronger co-occurrence, meaning that the corresponding challenges appear together in more Stack Overflow questions.}
    \label{fig:cooccurrence_heatmap_comparison}
\end{figure*}

Developer questions often contain more than one \gls{sst} challenge, so we analyze co-occurring subcategory pairs using raw co-occurrence count, Jaccard similarity, and lift. Figure~\ref{fig:cooccurrence_heatmap_comparison} visualizes the co-occurrence structure at category and subcategory levels, and Table~\ref{tab:subcategory_cooccurrence} reports the top co-occurring subcategory pairs across developer-reported security testing discussions.

\begin{table}[!t]
\centering
\caption{Top Co-occurring \gls{sst} Challenge Subcategories}
\label{tab:subcategory_cooccurrence}
\small
\setlength{\tabcolsep}{4pt}
\renewcommand{\arraystretch}{1.10}
\begin{tabularx}{\columnwidth}{
>{\raggedright\arraybackslash}X
>{\raggedright\arraybackslash}X
>{\centering\arraybackslash}p{0.10\columnwidth}
>{\centering\arraybackslash}p{0.13\columnwidth}
>{\centering\arraybackslash}p{0.10\columnwidth}
>{\centering\arraybackslash}p{0.18\columnwidth}
}
\toprule
\textbf{Subcategory A} & \textbf{Subcategory B} & \textbf{Count} & \textbf{Jaccard} & \textbf{Lift} & \textbf{Type} \\
\midrule
Explainability & False positives & 14 & 0.109 & 1.628 & Within-category \\
Integration & Tool suggestions & 9 & 0.086 & 1.645 & Cross-category \\
Documentation / Resources & Integration & 9 & 0.093 & 1.965 & Cross-category \\
False positives & Inconsistency & 5 & 0.049 & 1.219 & Within-category \\
Configuration & Integration & 5 & 0.044 & 0.845 & Cross-category \\
Configuration & Fuzzing setup & 5 & 0.094 & 10.981 & Within-category \\
Coverage & Explainability & 4 & 0.054 & 2.755 & Cross-category \\
Features & Tool suggestions & 4 & 0.056 & 1.827 & Within-category \\
Configuration & Documentation / Resources & 4 & 0.044 & 1.071 & Cross-category \\
Explainability & Reporting & 4 & 0.049 & 1.791 & Within-category \\
\bottomrule
\end{tabularx}
\end{table}

The co-occurrence results show that \textit{Explainability} and \textit{False positives} form the strongest pair, with 14 co-occurring questions. This pattern indicates that false positives are also interpretation problems, where developers need evidence to decide whether scanner findings are real vulnerabilities or false alarms. \textit{Integration} also co-occurs with \textit{Tool suggestions} and \textit{Documentation / Resources}, showing that tool adoption problems combine setup, selection, and guidance needs. Configuration-related pairs further show that setup issues can affect security testing reliability, especially when \textit{Configuration} co-occurs with \textit{Integration}, \textit{Documentation / Resources}, and \textit{Fuzzing setup}.

\begin{findingbox}
\textbf{Finding 14}. \textit{\gls{sst} challenges frequently co-occur across subcategories. The strongest pair is explainability and false positives, indicating that developers need clearer evidence to judge scanner findings. Integration also co-occurs with tool suggestions and documentation/resources, showing that tool adoption challenges often combine setup, selection, and guidance needs.}
\end{findingbox}

\section{Findings from RQ3: Taxonomy Stability on Temporally Unseen Questions}
\label{sec:rq3_taxonomy_stability}

This section reports the held-out stability analysis of the proposed \gls{sst} taxonomy. We use 21 temporally unseen \gls{sta} questions posted after August 2025 to examine whether newer developer questions can be mapped to the existing taxonomy. These questions were excluded from keyword refinement, inclusion/exclusion filtering, open coding, subcategory refinement, and category construction, making them an independent validation sample. All 21 held-out questions were mapped to at least one existing subcategory, and no question required a new subcategory or top-level category. Based on Equation~\ref{eq:heldout_coverage}, the taxonomy achieved 100\% held-out coverage.

The held-out questions span multiple taxonomy areas. The most frequent category is \textit{Security Testing Guidance and Tool Selection}, with 11 assignments, followed by \textit{Security Finding Interpretation and Reliability}, with six assignments, and \textit{Security Tool Integration and Automation}, with five assignments. Other assignments map to configuration, authentication testing, rule/policy/coverage validation, and security risks/constraints/performance categories. Four held-out questions contain multiple \gls{sst} concerns, such as intermittent 403 responses mapping to \textit{Authentication functionality testing}, \textit{Inconsistency}, and \textit{Configuration}. These mappings indicate that the taxonomy can explain newer and coupled developer-reported \gls{sst} challenges without structural changes.

\begin{findingbox}
\textbf{Finding 15}. \textit{The proposed \gls{sst} taxonomy remains stable on temporally unseen questions. All 21 held-out \gls{sta} questions posted after August 2025 were mapped to existing subcategories, resulting in 100\% coverage without requiring any new subcategory or top-level category.}
\end{findingbox}

\section{Discussion}
\label{sec:discussions}

In this section, we discuss the implications of our findings for \gls{sst} research, practice, tool support, and education.

\subsection{Improving Finding Interpretation and Remediation Support}

Our findings show that \textit{Security Finding Interpretation and Reliability} is the most discussed category and is increasing over time. The strong co-occurrence between \textit{False positives} and \textit{Explainability} indicates that developers need clearer evidence to decide whether scanner findings are exploitable issues or false alarms. Future research should develop explanation techniques that show why a finding was reported, which code or dependency path caused it, what evidence supports it, and whether it is reachable or exploitable in the application context~\cite{morshed2026use}. Similarly, the increasing trend in \textit{Remediation, Dependency, and Supply Chain Actionability} shows the need for techniques that connect dependency findings with provenance, transitive dependency paths, runtime reachability, exploitability evidence, and version-aware remediation options~\cite{ryan2025unveiling,haque2025security}.

\subsection{Supporting Workflow-Aware SST Integration and Validation}

Our results show that \gls{sst} tools are difficult to integrate into CI/CD pipelines, testing frameworks, browsers, IDEs, cloud platforms, containers, reporting systems, and authenticated environments. Because integration often co-occurs with tool suggestions and documentation/resources, future work should provide reusable pipeline patterns, standardized configuration templates, authenticated scan support, and interoperable reporting formats. The difficulty of \textit{Security Rule, Policy, and Coverage Validation} also suggests that developers need stronger \gls{sst}-specific coverage and validation models that capture vulnerability-class coverage, endpoint coverage, authentication-state coverage, rule coverage, dependency coverage, and policy-enforcement coverage.

\subsection{Strengthening Authentication Testing and Developer Assistance}

Authentication and authorization testing remains a distinct \gls{sst} challenge because protected functionality often depends on realistic users, roles, sessions, tokens, and external identity providers. Future research should investigate reusable authentication-aware \gls{sst} frameworks that support mock users, realistic security contexts, role-based access-control tests, token-based authentication, and external identity-provider integration. The held-out stability analysis also suggests that the taxonomy can support developer assistance systems. Such systems could classify new \gls{sst} questions, recommend documentation, identify likely solution strategies, connect related discussions, and support automated triage in forums, issue trackers, and tool reports.

\subsection{Implications for Practice, Tools, and Education}

Practitioners should treat \gls{sst} as an end-to-end workflow rather than a single scanning activity. Developers need support before testing, such as tool selection and configuration; during testing, such as integration and validation; and after testing, such as finding interpretation, false-positive assessment, vulnerability localization, and remediation. Tool providers should improve explanation quality, integration support, and remediation actionability through clearer reports, affected source locations, dependency paths, confidence information, and reusable workflow examples. Educators and community maintainers should organize \gls{sst} learning materials around practical workflows, including tool selection, scanner configuration, authenticated endpoint testing, false-positive handling, CI/CD integration, and vulnerability remediation.
\section{Threats to Validity}
\label{sec:threat_to_validity}
This section summarizes the main validity threats related to data retrieval, manual labeling, metric interpretation, and generalizability.

\subsection{Internal Validity}

Internal validity concerns whether our study design accurately captures developer-reported \gls{sst} challenges. Keyword-based retrieval may miss relevant \gls{sta} questions because developers use diverse terminology, tool names, abbreviations, and informal descriptions. It may also retrieve questions that mention security testing terms without describing an actual \gls{sst} challenge. We reduced this risk by deriving keywords from academic and practitioner-oriented sources, inspecting sampled results, removing overly generic terms, and applying systematic inclusion and exclusion criteria. Manual labeling may also introduce annotator judgment in category and subcategory assignments. We mitigated this threat through a coding guide, independent annotation, inter-rater agreement measurement, disagreement resolution, and iterative label refinement. Finally, \gls{sta} metadata used for prevalence and difficulty may be influenced by question age, wording quality, tag visibility, topic popularity, user reputation, or editing history. We therefore use multiple complementary metrics and interpret them together with manually analyzed question content.

\subsection{External Validity}

External validity concerns whether the findings generalize beyond our dataset. Because we analyze publicly available \gls{sta} questions, the results may not represent all \gls{sst} practices, especially in proprietary, regulated, or security-sensitive environments where developers discuss challenges in private channels, issue trackers, vendor portals, or incident-response systems. Therefore, categories related to organizational policy, confidential vulnerability handling, compliance-driven testing, and internal security review may be underrepresented. The reported frequencies should be interpreted as the distribution of developer-reported \gls{sst} challenges in our filtered \gls{sta} dataset, not as absolute distributions of all \gls{sst} work. We partially address this threat through held-out validation with temporally unseen questions, but broader generalization requires replication across GitHub issues, vendor forums, security tool communities, mailing lists, and industrial datasets.

\section{Conclusion and Future Work}
\label{sec:conclusion}

This paper presents an empirical study of developer-reported \gls{sst} challenges on \gls{sta}. From 17,743 collected questions, we identify 582 \gls{sst}-relevant questions and construct a hierarchical taxonomy with 8 categories and 31 subcategories. The results show that developers face challenges across the \gls{sst} workflow, including tool selection, integration, configuration, authentication testing, finding interpretation, dependency analysis, and remediation actionability. Authentication and authorization testing receives high visibility and stronger resolution support, while security rule, policy, and coverage validation requires more technical context and longer resolution time. Temporal analysis shows increasing attention to security finding interpretation and remediation-related actionability, and co-occurrence analysis shows that challenges often appear as connected problem clusters, such as false positives with explainability and integration with tool suggestions and documentation/resources. The held-out analysis further shows that all temporally unseen questions can be mapped to existing subcategories, providing initial evidence of taxonomy stability.

Future work will validate the taxonomy through surveys and interviews with software developers, security engineers, DevSecOps practitioners, and tool builders. We also plan to study additional developer discussion sources, such as GitHub issues, vendor forums, security tool communities, and industrial datasets, to improve the generalizability of the findings.

\bibliographystyle{elsarticle-num} 
\bibliography{citations}

\end{document}